\documentclass[12pt,a4paper]{article}
\usepackage{amssymb}
\usepackage{amsmath}
\newcommand{\A}[1]{A^{(#1)}}
\newcommand{\C}[1]{C^{(#1)}}
\newcommand{\G}[1]{G^{(#1)}}
\newcommand{\CHI}[1]{\chi^{(#1)}}
\newcommand{\GAM}[1]{\Gamma^{(#1)}}

\newcommand{\LAM}[1]{\Lambda^{(#1)}}
\newcommand{\PSI}[1]{\Psi^{(#1)}}
\newcommand{\ft}[2]{{\textstyle\frac{#1}{#2}}}
\newcommand{\Int}{\mathop{\rm Int}\nolimits}
\newcommand{\nuv}{\underline{\phantom{\mu }\!\!\!\nu}}
\newcommand{\wdg}{{\scriptscriptstyle \wedge}}
\def\rmi{{\rm i}}
\def\rmd{{\rm d}}
\newcommand{\tNS}{}   

\csname @addtoreset\endcsname{equation}{section}
\begin{document}
\begin{titlepage}
\begin{flushright}
  UG-00-15 \\
  SU-ITP-01/09 \\
  IFT-UAM/CSIC-00-35 \\
  KUL-TF-01/06 \\
  hep-th/0103233
\end{flushright}
\begin{center}
\vspace{.5cm}\baselineskip=16pt {\LARGE \bf New Formulations of $D=10$
 Supersymmetry \\

\

 and D$8$--O$8$ Domain Walls } \vskip 1 cm {\Large
  Eric Bergshoeff$^1$, Renata Kallosh$^2$, Tom\'as Ort\'{\i}n$^3$, \\[3mm]
Diederik Roest$^1$, and Antoine Van Proeyen$^4$
} \\
\vskip 1 cm
{\small
  $^1$ Institute for Theoretical Physics, Nijenborgh 4, 9747 AG Groningen, The Netherlands \\[3mm]
  $^2$ Department of Physics, Stanford University, Stanford, California 94305,
USA
\\[3mm]
  $^3$ Instituto de F\'{\i}sica Te\'orica, C-XVI, Universidad Aut\'onoma de Madrid, E-28049-Madrid, Spain \\
       I.M.A.F.F., C.S.I.C., Calle de Serrano 113 bis, E-28006-Madrid, Spain \\[3mm]
  $^4$ Instituut voor Theoretische Fysica, Katholieke Universiteit Leuven,\\
       Celestijnenlaan 200D B-3001 Leuven, Belgium
}
\end{center}
\centerline{ABSTRACT}
\bigskip
We discuss a generalized form of  IIA/IIB supergravity depending on {\it
all} R-R potentials $\C{p}\ (p = 0, 1, \ldots 9)$ as the effective field
theory of Type~IIA/IIB superstring theory. For the IIA case we explicitly
break this R-R democracy to either $p\leq3$ or $p\geq5$ which allows us
to write a new bulk action that can be coupled to $N=1$ supersymmetric
brane actions.
\par
The case of 8-branes is studied in detail using the new bulk \& brane
action. The supersymmetric negative tension branes without matter
excitations can be viewed as orientifolds in the effective action. These
D$8$-branes and O$8$-planes are fundamental in Type ${\rm I}^\prime$
string theory. A BPS 8-brane solution is given which satisfies the jump
conditions on the wall. It implies a quantization of the mass parameter
in string units. Also we find a maximal distance between the two walls,
depending on the string coupling and the mass parameter. We derive the
same results via supersymmetric flow equations.
\end{titlepage}


\section{Introduction}

The initial purpose of this work was to construct supersymmetric domain
walls of string theory in $D=10$ which may shed some light on the stringy
origin of the brane world scenarios. In the process of pursuing this goal
we have realized that all descriptions of the  effective field theory of
Type~IIA/B string theory available in the literature are inefficient for
our purpose. This has led us to introduce new versions of the effective
supergravities corresponding to Type~IIA/B string theory.

The standard IIA massless supergravity includes the $\C{1}$ and $\C{3}$
R-R potentials and the corresponding $\G{2}$ and  $\G{4}$ gauge-invariant
R-R forms. Type IIB supergravity includes the $\C{0}$, $ \C{2}$ and
$\C{4}$  R-R potentials and the corresponding $\G{1}$, $\G{3}$ and
(self-dual) $\G{5}$  gauge-invariant R-R forms. On the other hand, string
theory has all D$p$-branes, odd and even, including the exotic ones, like
8-branes in IIA and 7-branes in IIB theory. These branes, of co-dimension
1 and 2,  are special objects which are different in many respects from
the other BPS-extended objects like the $p$-branes with $0\le p \le 6$,
which have co-dimension greater than or equal to~3. The basic difference
is in the behavior of the form fields at large distance, $\G{p+2}\sim
r^{p-8}$. For example the $\G{10}$ R-R form of the 8-brane does not fall
off at infinity but takes a constant value there. It is believed that such
extended objects can not exist independently but only in connection with
orientifold planes~\cite{PolchinskiBook}. However, the realization of the
total system in supergravity is rather obscure.

It has been realized a while ago~\cite{Polchinski:1995mt}  that  massive
IIA supergravity, discovered by Romans~\cite{Ro86}, was the key to
understand the spacetime picture of the 8-branes, which are domain walls
in $D=10$. A significant progress towards the understanding of the
8-brane solutions was made in~\cite{Polchinski:1996df,Bergshoeff:1996ui},
where the bulk supergravity solution was found. Also,
in~\cite{Bergshoeff:1996ui}, the description of the cosmological constant
via a 9-form potential, based upon the work
of~\cite{Duff:1980qv,Aurilia:1980xj}, was discussed.
In~\cite{Alonso-Alberca:2000ne} a standard 8-brane action coupling to this
9-form potential has been shown to be the appropriate source for the
second Randall--Sundrum scenario~\cite{RS}. Solutions for the coupled bulk
\& brane action system automatically satisfy the jump conditions and so
they are consistent, at least from this point of view. A major unsolved
problem was to find an explicitly supersymmetric description of coupled
bulk \& brane systems like it was done in~\cite{Bergshoeff:2000zn}. Such
a description should allow us to find out some important properties of
the domain walls like the distance between the planes, the status of
unbroken supersymmetry in the bulk and on the brane  etc. We expect that
realizing such a bulk \& brane  construction will lead to a better
insight into the  fundamental nature of extended objects of string theory.
\par
The string backgrounds that we want to describe using an explicitly
supersymmetric bulk \& brane action are one-dimensional orbifolds
obtained by modding out the circle $S^{1}$ by a reflection
$\mathbb{Z}_{2}$. The orbifold direction is the transverse direction of
the branes that fill the rest of the spacetime. Now, the orbifold
$S^{1}/\mathbb{Z}_{2}$ being a compact space, we cannot place a single
charged object (a D$8$-brane, say) in it, but we have to have at least
two oppositely charged objects. However, this kind of system cannot be in
supersymmetric equilibrium unless their tensions also have opposite
signs. We are going to identify these negative-tension objects with
O$8$-planes and we will propose an O$8$-plane action to be coupled to the
bulk supergravity action. O$8$-planes can only sit at orbifold points
because they require the spacetime to be mirror symmetric in their
transverse direction and, thus, they can sit in any of the two endpoints
of the segment $S^{1}/\mathbb{Z}_{2}$. We are going to place the other
(positive tension, opposite R-R charge) brane at the other endpoint.
Clearly we can, from the effective action point of view, identify the
positive tension brane as a combination of O$8$-planes and D$8$-branes
with positive total tension and the negative tension brane as a
combination of O$8$-planes and D$8$-branes with negative total tension.
\par
Our strategy will be to generalize the 5-dimensional construction of the
supersymmetric bulk \& brane action, proposed
in~\cite{Bergshoeff:2000zn}. The construction of~\cite{Bergshoeff:2000zn}
allowed to find a supersymmetric realization of the brane-world scenario
of Randall and Sundrum~\cite{RS}. We will repeat the construction
of~\cite{Bergshoeff:2000zn} in $D=10$ with the aim to get a better
understanding of branes and planes in string theory.

To solve the discrepancy between the bulk actions with limited field
content (lower-rank R-R forms) and the wide range of brane actions that
involve all the possible R-R forms, we have constructed a new formulation
of IIA/IIB supergravity up to quartic order in fermions. In particular,
the new formulation gives an easy control over the exotic $\G{0}$ and
$\G{10}$ R-R forms associated with the mass and cosmological
constant\footnote{In this paper we will mean by
  cosmological constant the square of the mass parameter of the
  Type~IIA supergravity.  Strictly speaking it is not a cosmological
  constant because in the Einstein frame it carries a dilaton factor.}
of the $D=10$ supergravity. This in turn allows a clear study of the
D$8$--O$8$ system describing a pair of supersymmetric domain walls which
are fundamental objects of the Type~${\rm I^\prime}$ string theory. The
quantization of the mass parameter and cosmological constant in stringy
units are simple consequences of the theory. Apart from being a tool to
understand the supersymmetric domain walls we were interested in, it can
be expected that the new effective theories of $D=10$ supersymmetry will
have more general applications in the future.
\par
This paper is organized as follows. First, in section~\ref{ss:susybulk}
we discuss the new formulations of $D=10$ supersymmetry in the bulk. In
subsection~\ref{ss:Dformulation} we give a democratic formulation based
upon a pseudo-action along the lines of~\cite{Bergshoeff:1996sq}. This
formulation is the one that treats all R-R potentials in a unified way,
leading to the same equations of motion as have been encountered
basically in the coupling to D-branes in~\cite{Green:1996bh}. For the
supersymmetric case, this field content and equations of motion follow
also from the superspace formulation
of~\cite{Cederwall:1997ri,Bergshoeff:1997tu}. However, only a
pseudo-action is available, whose equations of motion are supplemented by
duality constraints. It allows for a unified IIA/IIB treatment and has no
Chern--Simons terms. The constraints do not follow from the action and
have to be imposed by hand. These generalize the self-duality condition
that relates the components of the five-form field strength $\G{5}$ of
type~IIB theory to a set of relations between all Hodge dual field
strengths.  In IIB this self-duality prevents the construction of a proper
action. The same is true in this formulation with a pseudo-action and
generalized self-duality equations, for both type~IIA and IIB. In the IIA
case, different field strengths are related by the constraints and hence
the R-R democracy can be broken without paying a price.  In the IIB
theory, things are more complicated. The self-duality of $G^{(5)}$
prevents a similar construction. To allow for a proper action one has to
resort to a non-covariant formulation~\cite{Perry:1997mk} or the
auxiliary fields of Pasti, Sorokin and Tonin~\cite{Pasti:1995tn}.
\par
In subsection~\ref{ss:action} we break the self-duality for type~IIA, and
in this way are able to obtain a supersymmetric proper action with
potentials $\C{p}$, $p=5,7,9$. The duality relations of the first
formulation also do not allow to obtain the generalization of the
mechanism of~\cite{Bergshoeff:2000zn}, involving the replacement of the
mass parameter $G^{(0)}$ by a field $G^{(0)}(x)$. Indeed, to do so we
need an independent nine-form auxiliary field. In the democratic
formulation, its field strength $G^{(10)}$ is related to $G^{(0)}$
itself. In the formulation of section~\ref{ss:action}, the breaking of
the self-duality, at the same time of allowing an action, also has the
setting to allow for a varying $G^{(0)}(x)$. We will also see that the
democratic formulation does not preserve a suitable $\mathbb{Z}_2$
symmetry that played an important role in the mechanism
of~\cite{Bergshoeff:2000zn}, in case that the mass parameter is non-zero.
Therefore, it is the formulation of section~\ref{ss:action} that we will
use for the bulk \& 8-brane construction. A first version of this
formulation was given in~\cite{Bergshoeff:1996ui}. In
subsection~\ref{Cformulation} we obtain the string-frame version of the
standard Romans' massive IIA supergravity in which the self-duality is
also broken but potentials $\C{p}$ with $p=1,3$ are kept.
\par
Next, in section~\ref{branesusy} we discuss the supersymmetry on the
brane. The mechanism is the same for all branes, but only the action of
one brane is added, such that this breaks the democracy. The
supersymmetric bulk \& brane construction for the D$8$--O$8$ system is
discussed in section~\ref{8branesusy}. We propose a supersymmetric action
for the O$8$-planes, which together with 32 D$8$-branes allows us to
reinterpret the set of positive--negative tension branes as a combination
of O$8$-planes with D$8$-branes. In section~\ref{8brane} we give the
8-brane solution and calculate the corresponding Killing spinors. As an
application of our results, we discuss in section~\ref{ss:quantMcosmo}
the quantization of the mass and cosmological constant parameter
corresponding to this system. Also critical distances are discussed. In
section~\ref{ss:BPSact} we give the BPS action for the domain wall and
investigate the supersymmetric flow equations. Finally, in
section~\ref{ss:summary} we give a summary of our results and a
discussion. There are two appendices. In appendix~\ref{app:Conventions}
we give our conventions. In appendix~\ref{app:Tduality} we discuss the
world-sheet T-duality between the different brane systems.

\section{Supersymmetry in the Bulk} \label{ss:susybulk}
The standard formulation of $D=10$ IIA
(massless~\cite{Huq:1985im,Giani:1984wc,Campbell:1984zc} and
massive~\cite{Ro86}) and IIB~\cite{Schwarz:1983qr,Howe:1984sr}
supergravity has the following field content
\begin{eqnarray}
{\rm IIA}&:&\hskip 1truecm  \left\{
  g_{\mu \nu},
  B_{\mu \nu},
  \phi,
  \C{1}_{\mu},
  \C{3}_{\mu \nu \rho},
  \psi_\mu,
  \lambda
  \right\} \,, \cr
{\rm IIB}&:&\hskip 1truecm  \left\{
  g_{\mu \nu},
  B_{\mu \nu},
  \phi,
  \C{0},
  \C{2}_{\mu \nu},
  \C{4}_{\mu\nu\rho\sigma},
  \psi_\mu,
  \lambda
  \right\} \, .
\label{convfc}
\end{eqnarray}
In the IIA case, the massive theory contains an additional mass parameter
$G^{(0)} = m$. In the IIB case, an extra self-duality condition is imposed
on the field strength of the four-form. It turns out that one can realize
the N=2 supersymmetry on the R-R gauge fields of higher rank as well.
These are usually incorporated via duality relations. To treat the R-R
potentials democratically we propose in the next subsection a new
formulation based upon a pseudo-action. This democratic formulation
describes the dynamics of the bulk supergravity in the most elegant way.
However, it turns out that this formulation is not well suited for our
purposes. For the IIA case, we therefore give a different formulation in
subsection~\ref{ss:action} where the constant mass parameter has been
replaced by a field. The relation of this formulation with the standard
IIA supergravity (with the above field content) will be discussed in
subsection~\ref{Cformulation}.
\subsection{The Democratic Formulation}
\label{ss:Dformulation} To explicitly introduce the democracy among the
R-R potentials we propose a pseudo-action containing all potentials. Of
course this enlarges the number of degrees of freedom. Since a $p$- and
an $(8-p)$-form potential carry the same number of degrees of freedom,
the introduction of the dual potentials doubles the R-R sector. Including
the highest potential $\C{9}$ in IIA does not alter this, since it carries
no degrees of freedom. This 9-form potential can be seen as the potential
dual to the constant mass parameter $G^{(0)} = m$. The doubling of number
of degrees of freedom will be taken care of by a constraint, relating the
lower- and higher-rank potentials. This new formulation of supersymmetry
is inspired by the bosonic construction of~\cite{Fukuma:1999jt}, and, in
the case of IIB supergravity, is related to the pseudo-action construction
of~\cite{Bergshoeff:1996sq}.
\par
A pseudo-action~\cite{Bergshoeff:1996sq} can be used as a mnemonic to
derive the equations of motion. It differs from a usual action in the
sense that not all equations of motion follow from varying the fields in
the pseudo-action. To obtain the complete set of equations of motion, an
additional constraint has to be substituted by hand into the set of
equations of motion that follow from the pseudo-action. The constraint
itself does not follow from the pseudo-action. The construction we
present here generalizes the pseudo-action construction
of~\cite{Fukuma:1999jt,Bergshoeff:1996sq} in the sense that our
construction (i) treats the IIA and IIB case in a unified way,
introducing all R-R potentials in the pseudo-action, and (ii) describes
also the massive IIA case via a 9-form potential $C^{(9)}$ and a constant
mass parameter $G^{(0)} = m$.
\par
Our pseudo-action has the extended field content
\begin{eqnarray}
{\rm IIA}&:&\hskip 1truecm  \left\{
  g_{\mu \nu},
  B_{\mu \nu},
  \phi,
  \C{1}_{\mu},
  \C{3}_{\mu \nu \rho},
  \C{5}_{\mu \cdots \rho},
  \C{7}_{\mu \cdots \rho},
  \C{9}_{\mu \cdots \rho},
  \psi_\mu,
  \lambda
  \right\} \, , \cr
{\rm IIB}&:&\hskip 1truecm  \left\{
  g_{\mu \nu},
  B_{\mu \nu},
  \phi,
  \C{0},
  \C{2}_{\mu \nu},
  \C{4}_{\mu \cdots \rho},
  \C{6}_{\mu \cdots \rho},
  \C{8}_{\mu \cdots \rho},
  \psi_\mu,
  \lambda
  \right\} \, .
\label{fcdemo}
\end{eqnarray}
It is understood that in the IIA case the fermions contain both
chiralities, while in the IIB case they satisfy
\begin{align}
  \Gamma _{11} \psi_\mu = \psi_\mu\,, \qquad \Gamma _{11} \lambda = -
  \lambda\,, \qquad
  \text{(IIB).}
\label{chirality}
\end{align}
In that case they are doublets, and we suppress the corresponding index.
The explicit form of the pseudo-action is given by\footnote {We partly
use form notation. For our conventions, see
appendix~\ref{app:Conventions}.}
\begin{align}
 S_{\text{Pseudo}} =
  & - \frac{1}{2\kappa_{10}^2}\int d^{10} x \sqrt{-g}
    \Big\{
    e^{-2\phi} \Big[
    R\big(\omega(e)\big) -4\big( \partial{\phi} \big)^{2}
    +\tfrac{1}{2} H \cdot H + \nonumber\\
 &  \hspace{1cm}  -2\partial^\mu\phi \CHI{1}_{{\mu}}
        + H \cdot \CHI{3}
   +2 \bar{{\psi}}_{{\mu}}{\Gamma}^{{\mu}{\nu}{\rho}}
    {\nabla}_{{\nu}}{\psi}_{{\rho}}
    -2 \bar{{\lambda}}{\Gamma}^{{\mu}}
    {\nabla}_{{\mu}}{\lambda}
    +4 \bar{{\lambda}} {\Gamma}^{{\mu}{\nu}}
    {\nabla}_{{\mu}}{\psi}_{{\nu}}
    \Big]+ \displaybreak[2] \notag \\
  & \hspace{1cm} + \sum_{n=0,1/2}^{5,9/2}
    \tfrac{1}{4} \G{2n} \cdot \G{2n}
    + \tfrac{1}{2} \G{2n} \cdot \PSI{2n} \Big\}+ \mbox{ quartic fermionic terms}\, .
    \label{IIABactiondemo}
\end{align}
It is understood that the summation in the above pseudo-action is over
integers ($n=0,1,\ldots ,5$) in the IIA case and over half-integers
($n=1/2,3/2,\ldots ,9/2$) in the IIB case. In the summation range we will
always first indicate the lowest value for the IIA case, before the one
for the IIB case. Furthermore,
\begin{equation}
\frac{1}{2\kappa_{10}^2}= \frac{g^2}{2\kappa^2}=\frac{2\pi}{(2\pi
\ell_s)^8}\,, \label{coupling}
\end{equation}
where $\kappa^2$ is the physical gravitational coupling, $g$ is  the
string coupling constant and $\ell_s= \sqrt {\alpha '}$ is the string
length. For notational convenience we group all potentials and field
strengths in the formal sums
\begin{align}
  {\bf G} = \, \sum_{n=0,1/2}^{5,9/2} \G{2n} \,, \hspace{1.5cm}
  {\bf C} = \, \sum_{n=1,1/2}^{5,9/2} \C{2n-1} \,. \label{formsums}
\end{align}
The bosonic field strengths are given by\footnote{There is an
  alternative basis for the R-R potentials that we can call
  ``$\mathbf{A}$-basis'' and can be useful in certain contexts. This
  basis is related to the ``$\mathbf{C}$-basis'' just defined by
  \begin{equation}
\label{eq:Abasis}
 \mathbf{A}=\mathbf{C}\wdg \mathbf{e}^{-B} \, ,
  \end{equation}
with the following gauge transformations
\begin{equation}
\delta \mathbf{A}=d\mathbf{\Lambda}  -G^{(0)}\Lambda^{\tNS}
-\mathbf{A} \wdg d\Lambda\, ,
\end{equation}
and in it, the above R-R field strengths are written as follows:
\begin{equation}
 \mathbf{G}=\left(d\mathbf{A}+G^{(0)}\right)\wdg \mathbf{e}^{B}\, .
\end{equation}
The two main properties of this basis are that in it, the R-R potentials
only appear in the field strengths through their derivatives (field
strengths) and the standard Wess-Zumino term of the D$p$-brane actions
does not contain the NS-NS 2-form $B$. In the Type~IIB action written in
this basis, the invariance under constant shifts of the R-R scalar
(axion) is manifest. It is the existence of this basis that makes the
Scherk-Schwarz generalized dimensional reduction of
Ref.~\cite{Bergshoeff:1996ui}.}
\begin{align}
  H = d B \,, \qquad
  {\bf G} = d {\bf C} - d B \wdg {\bf C} + \G{0} {\bf e}^B \,,
\label{G2n}
\end{align}
where it is understood that each equation involves only one term from the
formal sums (\ref{formsums}) (only the relevant combinations are
extracted). The corresponding Bianchi identities then read
\begin{align}
  d H =0 \,, \hspace{1cm}
  d {\bf G} - H \wdg {\bf G} =0 \,.
\label{Bianchis}
\end{align}
In this subsection $G^{(0)} =m$ indicates the constant mass parameter of
IIA supergravity. In the IIB theory all equations should be read with
vanishing $\G{0}$. The spin connection in the covariant derivative
$\nabla_\mu $ is given by its zehnbein part: $\omega_{\mu}^{\; \;
ab}=\omega_{\mu}^{\; \; ab}(e)$. The bosonic fields couple to the
fermions via the bilinears $\CHI{1,3}$ and $\PSI{2n}$, which read
\begin{align}
  \CHI{1}_\mu =
  & -2 \bar{\psi}_\nu \Gamma^\nu \psi_\mu
    -2 \bar{\lambda} \Gamma^\nu \Gamma_\mu \psi_\nu\, , \notag \\
  \CHI{3}_{\mu\nu\rho} =
  & \tfrac{1}{2}\bar{{\psi}}_{{\alpha}}
    {\Gamma}^{[{\alpha}}
    \Gamma_{\mu\nu\rho}
    {\Gamma}^{{\beta}]}
    {\cal P}{\psi}_{{\beta}}
    + \bar{{\lambda}}
    \Gamma_{\mu\nu\rho}{}^{\beta}
    {{\cal P}}{\psi}_{{\beta}}
    -\tfrac{1}{2}\bar{{\lambda}} {\cal P}\Gamma_{\mu\nu\rho}
    {\lambda}\, , \notag \\
  \PSI{2n}_{\mu_1\cdots \mu_{2n}} =
  & {\textstyle\frac{1}{2}}e^{-{\phi}}
    \bar{{\psi}}_{{\alpha}}
    {\Gamma}^{[{\alpha}}
    \Gamma_{\mu_1\cdots \mu_{2n}}
    {\Gamma}^{{\beta}]}
    {\cal P}_n {\psi}_{{\beta}}
    +{\textstyle\frac{1}{2}}e^{-{\phi}}
    \bar{{\lambda}}
    \Gamma_{\mu_1\cdots \mu_{2n}}
    {\Gamma}^{{\beta}}
    {\cal P}_n{\psi}_{{\beta}}+ \notag \\
  & -{\textstyle\frac{1}{4}}
    e^{-{\phi}}
    \bar{{\lambda}}
    \Gamma_{ [ \mu_1\cdots \mu_{2n-1}}
    {\cal P}_n \Gamma_{\mu_{2n} ] } {\lambda}\, .
\label{fermionbilinears}
\end{align}
We have used the following definitions:
\begin{eqnarray}
{\cal P} &=& \Gamma_{11}\ \ \quad {\rm (IIA)} \qquad {\rm or}\qquad   -\sigma^3\ {\rm (IIB)}\,, \nonumber\\
{\cal P}_n &=& (\Gamma_{11})^{n}\ {\rm (IIA)}\qquad  {\rm or}\qquad
\sigma^1\ ({\rm n+1/2\ even}),\ \rmi\sigma^2\ ({\rm n+1/2\ odd}) \ ({\rm
IIB})\,.
\end{eqnarray}
Note that the fermions satisfy
\begin{equation}
\PSI{2n} = (-)^{\Int[n]+1} \star \PSI{10-2n}\,,
\label{Psidual}
\end{equation}
due to the $\Gamma$-matrices identity (\ref{starGamma}).
\par
Due to the appearance of all R-R potentials, the number of degrees of
freedom in the R-R sector has been doubled. Each R-R potential leads to a
corresponding equation of motion:
\begin{align}
 d \star (\G{2n}+\PSI{2n})
  + H \wdg \star (\G{2n+2}+\PSI{2n+2}) = 0\, . \label{eoms}
\end{align}
Now, one must relate the different potentials to get the correct number of
degrees of freedom. We therefore by hand impose the following duality
relations
\begin{align}
  \G{2n} + \PSI{2n} = (-)^{\Int[n]} \star \G{10-2n}\, ,
\label{duality}
\end{align}
in the equations of motion that follow from the pseudo-action
(\ref{IIABactiondemo}). It is in this sense that the action
(\ref{IIABactiondemo}) cannot be considered as a true action. Instead, it
should be considered as a mnemonic to obtain the full equations of motion
of the theory. As usual, the Bianchi identities and equations of motions
of the dual potentials correspond to each other when employing the
duality relation. For the above reason the democratic formulation can be
viewed as self-dual, since (\ref{duality}) places constraints relating
the field content (\ref{fcdemo}).
\par
The pseudo-action (\ref{IIABactiondemo}) is invariant under supersymmetry
provided we impose the duality relations (\ref{duality}) after varying
the action. The supersymmetry rules read (here given modulo cubic fermion
terms):
\begin{align}
  \delta_{{\epsilon}} {e}_{{\mu}}{}^{{a}} =
  & \bar{{\epsilon}}{\Gamma}^{{a}} {\psi}_{{\mu}}\, , \notag \\
  \delta_{{\epsilon}} {\psi}_{{\mu}} =
  & \Big( \partial_{{\mu}} +\tfrac{1}{4}
    \not\!{\omega}_{{\mu}}
    +\tfrac{1}{8}{\cal P}\not\!\! {H}_{\mu}
    \Big) {\epsilon}
    +\tfrac{1}{16} e^{{\phi}} \sum_{n=0,1/2}^{5,9/2} \frac{1}{(2n)!}
    \not \! {G}^{(2n)} {\Gamma}_{{\mu}}
    {\cal P}_n{\epsilon}\, , \notag \\
  \delta_{{\epsilon}} B_{\mu\nu} =
  & -2 \, \bar{{\epsilon}} \Gamma_{[\mu}
    {\cal P} {\psi}_{\nu]}\, , \notag \\
  \delta_\epsilon \C{2n-1}_{\mu_1\cdots \mu_{2n-1}} =
   & - e^{-\phi} \, \bar{\epsilon} \,
    \Gamma_{[\mu_1\cdots \mu_{2n-2}} \, {\cal P}_n \,
\Big((2n-1) \psi_{\mu_{2n-1}]} - \tfrac{1}{2} \Gamma_{\mu_{2n-1}]}
\lambda\Big)
    + \notag \\
    & +(n-1)(2n-1) \, \C{2n-3}_{[\mu_1\cdots \mu_{2n-3}} \,
   \delta_\epsilon B_{\mu_{2n-2}\mu_{2n-1}]}\, , \displaybreak[2] \notag \\
  \delta_{{\epsilon}}{\lambda} =
  & \Big( \! \! \not \! \partial \phi
    + \tfrac{1}{12} \not\!\! {H} {\cal P}\Big) {\epsilon}
    + \tfrac{1}{8} e^{{\phi}} \sum_{n=0,1/2}^{5,9/2} (-)^{2n} \frac{5-2n}{(2n)!}
   \not\! {G}^{(2n)} {\cal P}_n
    {\epsilon}\, , \notag \\
  \delta_{{\epsilon}}{\phi} =
  & \tfrac{1}{2} \, \bar{{\epsilon}}{\lambda}\,,
\label{IIABsusydemo}
\end{align}
where $\epsilon $ is a spinor similar to $\psi _\mu $, i.e. in IIB:
$\Gamma _{11}\epsilon =\epsilon$. Note that for $n$ half-integer (the IIB
case) these supersymmetry rules exactly reproduce the rules given in
eq.~(1.1) of~\cite{BRJO99}.
\par
Secondly, the pseudo-action (\ref{IIABactiondemo}) is also invariant under
the usual bosonic NS-NS and R-R gauge symmetries with parameters
$\Lambda^{\tNS}$ and $\LAM{2n}$ respectively:
\begin{align}
  \delta_\Lambda B =  \, d \Lambda^{\tNS} \,, \qquad
  \delta_\Lambda {\bf C} = \, (d {\bf \Lambda}
     - \G{0} \Lambda^{\tNS}) \wdg {\bf e}^B \,, \qquad
  \text{with~~} {\bf \Lambda} = \sum_{n=0,1/2}^{4,7/2} \LAM{2n} \,.
\label{gaugedemo}
\end{align}
\par
Finally, there is a number of $\mathbb{Z}_2$-symmetries. However, in the
IIA case these $\mathbb{Z}_2$-symmetries are {\it only valid for $G^{(0)}
= m =0$}.  The world-sheet form of these symmetries is given in
appendix~\ref{app:Tduality}. Below we show how these symmetries of the
action act on supergravity fields. For both massless IIA and IIB there is
a fermion number symmetry $(-)^{F_L}$ given by
\begin{align}
  \big\{ \phi, g_{\mu \nu}, B_{\mu \nu} \big\}
  & \rightarrow \big\{ \phi, g_{\mu \nu}, B_{\mu \nu} \big\}\, , \notag \\
  \big\{\C{2n-1}_{\mu_1 \cdots \mu_{2n-1}} \big\}
  & \rightarrow - \big\{ \C{2n-1}_{\mu_1 \cdots \mu_{2n-1}} \big\} \, , \notag \\
  \big\{ \psi_\mu , \lambda, \epsilon \big\}
  & \rightarrow  + {\cal P}
    \big\{ \psi_\mu , - \lambda, \epsilon \big\}\, , \text{~~(IIA),} \notag \\
  \big\{ \psi_\mu , \lambda, \epsilon \big\}
  & \rightarrow  + {\cal P}
    \big\{ \psi_\mu , \lambda, \epsilon \big\}\, , \text{~~(IIB).}
\label{Omegahat}
\end{align}
In the IIB case there is an additional worldsheet parity symmetry
$\Omega$ given by
\begin{align}
  \big\{ \phi, g_{\mu \nu}, B_{\mu \nu} \big\}
  & \rightarrow \big\{ \phi, g_{\mu \nu}, - B_{\mu \nu} \big\}\, , \notag \\
  \big\{\C{2n-1}_{\mu_1 \cdots \mu_{2n-1}} \big\}
  & \rightarrow (-)^{n+1/2}\big\{ \C{2n-1}_{\mu_1 \cdots \mu_{2n-1}} \big\} \, , \notag \\
  \big\{ \psi_\mu , \lambda, \epsilon \big\}
  & \rightarrow \sigma^1
    \big\{ \psi_\mu , \lambda, \epsilon \big\}\, ,
\label{Omega}
\end{align}
In the massless IIA case there is a similar $I_9\Omega$-symmetry involving
an additional parity transformation in the 9-direction. Writing $\mu =
(\underline \mu, \dot 9)$, the rules are given by
\begin{align}
  x^{\dot 9} & \rightarrow -x^{\dot 9}\, , \notag \\
  \big\{ \phi, g_{\underline{\mu} \nuv },
     B_{\underline{\mu} \nuv} \big\}
  & \rightarrow \big\{ \phi, g_{\underline{\mu} \nuv},
    -B_{\underline{\mu} \nuv} \big\}\, , \notag \\
  \big\{ \C{2n-1}_{\underline{\mu_1} \cdots \underline{\mu_{2n-1}}} \big\}
  & \rightarrow (-)^{n+1} \big\{
    \C{2n-1}_{\underline{\mu_1} \cdots \underline{\mu_{2n-1}}} \big\}
    \,, \notag \\
  \big\{ \psi_{\underline{\mu}} , \lambda, \epsilon \big\}
  & \rightarrow +  \Gamma^{9} \big\{ \psi_{\underline{\mu}} ,
   -\lambda, \epsilon \big\}\,.
\label{I9Omega}
\end{align}
The parity of the fields with one or more indices in the
$\dot{9}$-direction is given by the rule that every index in the
$\dot{9}$-direction gives an extra minus sign compared to the above rules.
\par
In both IIA and IIB there is also the obvious symmetry of interchanging
all fermions by minus the fermions, leaving the bosons invariant.
\par
The $\mathbb{Z}_2$-symmetries are used for the construction of superstring
theories with sixteen supercharges, see~\cite{Bergshoeff:2000re}.
$(-)^{F_L}$ gives a projection to the $E_8 \times E_8$ heterotic
superstring (IIA) or the SO(32) heterotic superstring theory (IIB).
$\Omega$ is used to reduce the IIB theory to the SO(32) Type I
superstring, while the $I_9\Omega$-symmetry reduces the IIA theory to the
${\rm Type\ I}^\prime\, SO(16) \times SO(16)$ superstring theory.
\par
One might wonder at the advantages of the generalized pseudo-action
(\ref{IIABactiondemo}) above the standard supergravity formulation. At the
cost of an extra duality relation we were able to realize the R-R
democracy in the action. Note that only kinetic terms are present; by
allowing for a larger field content the Chern--Simons term is eliminated.
Under T-duality all kinetic terms are easily seen to transform into each
other~\cite{Meessen:1998qm}. The same goes for the duality constraints.
This formulation is elegant and comprises all potentials. However, it is
impossible to construct a proper action in this formulation due to the
doubling of the degrees of freedom. Therefore, to add brane actions to
the bulk system, the democratic formulation is not suitable. This is due
to two reasons. First, the $I_9\Omega$ symmetry is only valid for
$G^{(0)} = 0$, but we will need this symmetry in our construction of the
bulk \& 8-brane system. Secondly, to describe a charged domain wall, we
would like to have opposite values for $G^{(0)}$ at the two sides of the
domain wall, i.e.~we want to allow for a mass parameter that is only
piecewise constant. The R-R democracy has to be broken to accommodate for
an action and this will be discussed in the following two subsections.
\subsection{The Dual Formulation of IIA} \label{ss:action}

We will present here the new dual formulation with action, available for
the IIA case only. A proper action will be constructed in this
formulation. It is this formulation that we will apply in our
construction of the bulk \& brane system. We will call this the dual
formulation and explain why in the next subsection.
\par
The independent fields in this formulation are
\begin{equation}
  \left\{   e_\mu ^a,
  B_{\mu \nu},
  \phi,
  \G{0},
  \G{2}_{\mu \nu },
  \G{4}_{\mu _1\cdots \mu _4},
  \A{5}_{\mu _1\cdots\mu _5 },
  \A{7}_{\mu _1\cdots\mu _7 },
  \A{9}_{\mu _1\cdots\mu _9 },
  \psi_\mu,
  \lambda
  \right\}.
 \label{indepFields}
\end{equation}
The bulk action reads
\begin{align}
 S_{\text{bulk}} =
  & - \frac{1}{2\kappa_{10}^2}\int d^{10} x \sqrt{-g}
    \Big\{
    e^{-2\phi} \big[
    R\big(\omega(e)\big) -4\big( \partial{\phi} \big)^{2}
    +\tfrac{1}{2} H \cdot H
    -2\partial^{{\mu}}{\phi} \CHI{1}_{{\mu}}
        + H \cdot \CHI{3}+ \notag \\
  & +2 \bar{{\psi}}_{{\mu}}{\Gamma}^{{\mu}{\nu}{\rho}}
    {\nabla}_{{\nu}}{\psi}_{{\rho}}
    -2 \bar{{\lambda}}{\Gamma}^{{\mu}}
    {\nabla}_{{\mu}}{\lambda}
    +4 \bar{{\lambda}} {\Gamma}^{{\mu}{\nu}}
    {\nabla}_{{\mu}}{\psi}_{{\nu}}
    \big]
    +   \sum_{n=0,1,2}
     \tfrac{1}{2} \G{2n} \cdot \G{2n}
    + \G{2n} \cdot \PSI{2n}  +  \notag \\
  & - \star \, \big[
      \tfrac{1}{2} \, \G{4} \G{4} B
    - \tfrac{1}{2} \, \G{2} \G{4} B^2
    + \tfrac{1}{6} \, \G{2}{}^2 B^3
    + \tfrac{1}{6} \, \G{0} \G{4} B^3 -\tfrac{1}{8} \, \G{0} \G{2} B^4+ \notag \\
  & +\tfrac{1}{40} \, \G{0}{}^2 B^5
    + {\bf e}^{- B} {\bf G}
    d (\A{5} - \A{7} + \A{9}) \big] \Big\}  +\mbox{ quartic fermionic terms}\,,
    \label{dualaction}
\end{align}
where all $\wdg$'s have been omitted in the last two lines. In the last
term a projection on the 10-form is understood.  Here $\textbf{G}$ is
defined as in (\ref{formsums}) but where $\G{0}$, $\G{2}$ and $\G{4}$ are
now independent fields (which we will call black boxes) and are no longer
given by (\ref{G2n}). Note that their Bianchi identities are imposed by
the Lagrange multipliers $\A{9}$, $\A{7}$ and $\A{5}$. The NS-NS
three-form field strength is given by (\ref{G2n}).
\par
The symmetries of the action are similar to those of the democratic
formulation with some small changes. In the supersymmetry transformations
of gravitino and gaugino, the sums now extend only over $n=0,1,2$:
\begin{align}
  \delta_{{\epsilon}} {e}_{{\mu}}{}^{{a}} =
  & \bar{{\epsilon}}{\Gamma}^{{a}} {\psi}_{{\mu}}\, , \notag \\
  \delta_{{\epsilon}} {\psi}_{{\mu}} =
  & \Big( \partial_{{\mu}} +\tfrac{1}{4}
    \not\!{\omega}_{{\mu}}
    +\tfrac{1}{8}\Gamma_{11}\not\!\! {H}_{\mu}
    \Big) {\epsilon}
    +\tfrac{1}{8} e^{{\phi}} \sum_{n=0,1,2} \frac{1}{(2n)!}
    \not \! {G}^{(2n)} {\Gamma}_{{\mu}}
    ({\Gamma}_{11})^{n}{\epsilon}\, , \notag \\
  \delta_{{\epsilon}} B_{\mu\nu} =
  & -2 \, \bar{{\epsilon}} \Gamma_{[\mu}
    \Gamma_{11} {\psi}_{\nu]}\, , \notag \\
     \delta_{{\epsilon}}{\lambda} =
  & \Big( \! \! \not \! \partial \phi
    - \tfrac{1}{12} {\Gamma}_{11} \not\!\! {H} \Big) {\epsilon}
    + \tfrac{1}{4} e^{{\phi}} \sum_{n=0,1,2} \frac{5-2n}{(2n)!}
    \not\! {G}^{(2n)} ({\Gamma}_{11})^{n}
    {\epsilon}\,, \nonumber\\
  \delta_{{\epsilon}}{\phi} =
  & \tfrac{1}{2} \, \bar{{\epsilon}}{\lambda} \,, \nonumber\\
  \delta_\epsilon \textbf{A} =& \textbf{e}^{-B} \wdg \textbf{E}
  \,, \nonumber \\
\delta _{\epsilon} \textbf{G}= & d \textbf{E} + \textbf{G} \wdg \delta
_\epsilon B - H \wdg \textbf{E}
\,,   \notag \\
  \text{with~~}
  & E^{(2n-1)}_{\mu_1 \cdots \mu_{2n-1}} \equiv
    - e^{-\phi} \, \bar{\epsilon} \,
    \Gamma_{[\mu_1\cdots \mu_{2n-2}} \, (\Gamma_{11})^n \,
\Big((2n-1)\psi_{\mu_{2n-1}]} - \tfrac{1}{2} \Gamma_{\mu_{2n-1}]}
\lambda\Big)\,. \label{IIAsusydual}
\end{align}
The transformation of the black boxes $\textbf{G}$ follow from the
requirement that ${\bf e}^{-B} {\bf G}$ transforms in a total derivative.
Here the formal sums
\begin{align}
  {\bf A} = \sum_{n=1}^5 \A{2n-1}\,, \qquad
  {\bf E} = \sum_{n=1}^5 E^{(2n-1)}\,, \qquad
  {\bf G} =\sum_{n=0}^5 G^{(2n)}\,,
\label{AEsum}
\end{align}
have been used. Note that the first formal sum in (\ref{AEsum}) contains
fields, $A^{(1)}$ and $A^{(3)}$, that do not occur in the action. The same
applies to ${\bf G}$, which contains the extra fields $G^{(6)}, G^{(8)}$
and $G^{(10)}$. Although these fields do not occur in the action, one can
nevertheless show that the supersymmetry algebra is realized on them. To
do so one must use the supersymmetry rules of (\ref{IIAsusydual}) and the
equations of motion that follow from the action (\ref{dualaction}).
\par
The gauge symmetries with parameters $\Lambda^{\tNS}$ and $\LAM{2n}$ are
\begin{align}
  \delta_\Lambda B = & \, d \Lambda^{\tNS} \,, \qquad
    \delta_\Lambda {\bf A} =  \, d {\bf \Lambda} - \G{0}
\Lambda^{\tNS}
    - d \Lambda^{\tNS} \wdg \textbf{A} \,, \nonumber \\
  \delta_\Lambda {\bf G} = & \, d \Lambda \wdg
  \big({\bf G} - {\bf e}^B \wdg (d {\bf A} +\G{0}) \big)
  + {\bf e}^B \wdg \Lambda \wdg d \G{0} \,.
\label{gaugeA}
\end{align}
Note that, with respect to the R-R gauge symmetry, the ${\bf A}$
potentials transform as a total derivative while the black boxes are
invariant.
\par
Finally, there are $\mathbb{Z}_2$-symmetries, $(-)^{F_L}$ and $I_9\Omega$,
which leave the action invariant. In contrast to the democratic
formulation these two $\mathbb{Z}_2$-symmetries are valid symmetries even
for $G^{(0)}\ne 0$. The $(-)^{F_L}$-symmetry is given by
\begin{align}
  \big\{ \phi, g_{\mu \nu}, B_{\mu \nu} \big\}
  & \rightarrow \big\{ \phi, g_{\mu \nu}, B_{\mu \nu} \big\}\, , \notag \\
  \big\{ \G{2n}_{\mu_1 \cdots \mu_{2n}}, \A{2n-1}_{\mu_1 \cdots \mu_{2n-1}} \big\}
  & \rightarrow - \big\{ \G{2n}_{\mu_1 \cdots \mu_{2n}}, \A{2n-1}_{\mu_1 \cdots \mu_{2n-1}} \big\} \, , \notag \\
  \big\{ \psi_\mu , \lambda, \epsilon \big\}
  & \rightarrow  + \Gamma_{11}
    \big\{ \psi_\mu , -\lambda, \epsilon \big\}\,,
\end{align}
while the second $I_9\Omega$-symmetry reads
\begin{align}
  x^9 & \rightarrow -x^9\, , \notag \\
  \big\{ \phi, g_{\underline{\mu} \nuv},
   B_{\underline{\mu} \nuv} \big\}
  & \rightarrow \big\{ \phi, g_{\underline{\mu} \nuv},
    -B_{\underline{\mu} \nuv} \big\}\, , \notag \\
  \big\{ \G{2n}_{\underline{\mu_1} \cdots \underline{\mu_{2n}}},
    \A{2n-1}_{\underline{\mu_1} \cdots \underline{\mu_{2n-1}}} \big\}
  & \rightarrow (-)^{n+1} \big\{
    \G{2n}_{\underline{\mu_1} \cdots \underline{\mu_{2n}}},
    \A{2n-1}_{\underline{\mu_1} \cdots \underline{\mu_{2n-1}}} \big\}\,,
    \notag \\
  \big\{ \psi_{\underline{\mu}} , \lambda, \epsilon \big\}
  & \rightarrow +  \Gamma^{9} \big\{ \psi_{\underline{\mu}} ,
    -\lambda, \epsilon \big\}\, .
\end{align}

\subsection{The Standard Formulation of IIA}
\label{Cformulation}

Our dual action (\ref{dualaction}) can be reduced to the string-frame
version of the standard formulation of massive IIA supergravity
(originally written in Einstein frame in~\cite{Ro86}).  Here we will show
how to go from one formulation to the other.

Consider the field equations for the $\A 5$, $\A 7$ and $\A 9$ potentials
given by
\begin{align}
  d ( {\bf e}^{-B} \wdg {\bf G} ) =0 \,.
\end{align}
The most general solutions can be taken in the form
\begin{align}
 {\bf e}^{-B} \wdg {\bf G} = d {\bf A} + {\bf G}_{\rm flux}\,,
 \end{align}
or, explicitly,
\begin{align}
  & \G{0} = \G{0}_{\rm flux}\,, \qquad \G{2} = d \A{1} + \G{0} B + \G{2}_{\rm flux}
  \,, \notag \\
  & \G{4} = d \A{3} + \G{2} \wdg B - \tfrac{1}{2} \G{0} B \wdg B
  + \G{4}_{\rm flux} \,.
\label{solG}
\end{align}
The ${\bf G}_{\rm flux}$ are cohomological solutions. If there is full
10-dimensional Lorentz symmetry, then only $\G{0}_{\rm flux}$ can be
non-zero and is a constant, which is the mass parameter $m$ in the theory
of Romans. We will mostly consider this situation. However, before
proceeding, we can remark that we could consider that constant fluxes
$\G{2}_{\rm flux}$, $\G{4}_{\rm flux}$ are present in our configuration
in addition to $m$ and only 4-dimensional Lorentz symmetry is preserved
(e.g.~\cite{Polchinski:1996sm,Gukov:1999gr}).  See
section~\ref{ss:summary} for more comments.
\par
{}From now on we restrict ourselves to
\begin{equation}
\G{0}_{\rm flux}=m\,, \qquad \G{2}_{\rm flux}= \G{4}_{\rm flux}=0\,.
 \label{Gfluxism}
\end{equation}
Substituting these solutions in the bulk action \eqref{dualaction}, we
obtain a theory without black boxes but with a mass parameter and a one-
and a three- form. Note that this is the same field content
\eqref{convfc} as the massive theory of Romans. A close inspection of the
action reveals that
in fact, we are dealing with the standard Romans' theory
written in the $\mathbf{A}$-basis introduced in~(\ref{eq:Abasis}).

Note that, as we have already remarked, $\mathbf{A}= {\bf C}\wdg{\bf
  e}^{-B} $ is exactly the combination that occurs in the Wess--Zumino
terms of D-brane actions, with the world-volume fields put to zero.

In the $\bf C$-basis, the standard formulation has the action
\begin{align}
 S_{\text{bulk}} =
  & - \frac{1}{2\kappa_{10}^2}\int d^{10} x \sqrt{-g}
    \Big\{
    e^{-2\phi} \big[
    R\big(\omega(e)\big) -4\big( \partial{\phi} \big)^{2}
    +\tfrac{1}{2} H \cdot H
    -2\partial^{{\mu}}{\phi} \CHI{1}_{{\mu}}
    + H \cdot \CHI{3}+ \notag \\
  & +2 \bar{{\psi}}_{{\mu}}{\Gamma}^{{\mu}{\nu}{\rho}}
    {\nabla}_{{\nu}}{\psi}_{{\rho}}
    -2 \bar{{\lambda}}{\Gamma}^{{\mu}}
    {\nabla}_{{\mu}}{\lambda}
    +4 \bar{{\lambda}} {\Gamma}^{{\mu}{\nu}}
    {\nabla}_{{\mu}}{\psi}_{{\nu}}
    \big]
    + \sum_{n=0,1,2}
    \tfrac{1}{2} \G{2n} \cdot \G{2n}
    + \G{2n} \cdot \PSI{2n} + \notag \\
  & - \star \, \big[
    \tfrac{1}{2} \, d \C{3} \, d \C{3} \, B
    + \tfrac{1}{6} \, \G{0} \, d \C{3} \, B^3
    + \tfrac{1}{40} \, \G{0}{}^2 \, B^5 \big] \Big\}
    +\mbox{ quartic fermionic terms}\,,
\label{standaction}
\end{align}
where all $\wdg$'s have been omitted in the last line. Here all field
strengths $\G{2n}$ are given by \eqref{G2n} and $\G{0}$ is constant. Note
that the six Chern-Simons terms in \eqref{dualaction} can be written in
only three terms when the $\G{2n}$'s are field strengths. The standard
IIA action is invariant under the $N=2$ supersymmetry rules
\begin{align}
  \delta_{{\epsilon}} {e}_{{\mu}}{}^{{a}} =
  & \bar{{\epsilon}}{\Gamma}^{{a}} {\psi}_{{\mu}}\, , \notag \\
  \delta_{{\epsilon}} {\psi}_{{\mu}} =
  & \Big( \partial_{{\mu}} +\tfrac{1}{4}
    \not\!{\omega}_{{\mu}}
    +\tfrac{1}{8} \Gamma_{11} \not\!\! {H}_{\mu}
    \Big) {\epsilon}
    +\tfrac{1}{8} e^{{\phi}} \sum_{n=0,1,2} \frac{1}{(2n)!}
    \not \! {G}^{(2n)} {\Gamma}_{{\mu}}
    (\Gamma_{11})^n \, {\epsilon}\, , \notag \\
  \delta_{{\epsilon}} B_{\mu\nu} =
  & -2 \, \bar{{\epsilon}} \Gamma_{[\mu}
    \Gamma_{11} {\psi}_{\nu]}\, , \notag \\
  \delta_\epsilon \C{2n-1}_{\mu_1\cdots \mu_{2n-1}} =
   & - e^{-\phi} \, \bar{\epsilon} \,
     \Gamma_{[\mu_1\cdots \mu_{2n-2}} \, (\Gamma_{11})^n \,
     \Big((2n-1) \psi_{\mu_{2n-1}]} - \tfrac{1}{2} \Gamma_{\mu_{2n-1}]}
     \lambda\Big) + \notag \\
    & +(n-1)(2n-1) \, \C{2n-3}_{[\mu_1\cdots \mu_{2n-3}} \,
   \delta_\epsilon B_{\mu_{2n-2}\mu_{2n-1}]}\, , \displaybreak[2] \notag \\
  \delta_{{\epsilon}}{\lambda} =
  & \Big( \! \! \not \! \partial \phi
    + \tfrac{1}{12} \not\!\! {H} \Gamma_{11} \Big) {\epsilon}
    + \tfrac{1}{4} e^{{\phi}} \sum_{n=0,1,2} \frac{5-2n}{(2n)!}
   \not\! {G}^{(2n)} (\Gamma_{11})^n
    {\epsilon}\, , \notag \\
  \delta_{{\epsilon}}{\phi} =
  & \tfrac{1}{2} \, \bar{{\epsilon}}{\lambda}\,,
\label{IIAsusystand}
\end{align}
and the gauge transformations \eqref{gaugedemo}. Also the
$\mathbb{Z}_2$-symmetries \eqref{Omegahat} and \eqref{I9Omega} are valid
but {\it only for vanishing mass}.

The dual formulation of section~\ref{ss:action} thus can be converted to
the standard formulation. But it is also possible to revert this: Romans'
theory can be used to derive \eqref{dualaction}. To do so, one must
perform the first step in the dualization of the one- and three-form.
That is, the field strengths are promoted to black boxes and their
Bianchi identities are imposed by Lagrange multipliers, considering
$\G{0}$ as a zero-form field strength. The full dualization then implies
the elimination of the black boxes by solving their equations of motion.
Performing only the first step one obtains the action \eqref{dualaction}.
It is for this reason that we call it the dual formulation.

Thus we have three different formulations of one theory at our disposal.
The democratic formulation comprises all potentials but can not be used
to add brane actions since it has no proper action.  The standard
formulation does have a proper action, containing $\C{1}$ and $\C{3}$,
and thus is suitable for the $0$- and $2$-branes.  The dual action has
the dual potentials $\A{5}$, $\A{7}$ and $\A{9}$, accommodating for the
$4$-, $6$- and $8$-branes.  With the latter two formulations we have set
the stage for the addition of brane actions of any (even) dimension as
long as we do not add simultaneously higher- and lower-dimensional branes.

\section{Supersymmetry on the Brane}
\label{branesusy}  Having established supersymmetry in the bulk, we now
turn to supersymmetry on the brane. As mentioned in the introduction, our
main interest is in one-dimensional orbifold constructions with 8-branes
at the orbifold points. Using the techniques of the three-brane on the
orbifold in five dimensions~\cite{Bergshoeff:2000zn}, we want to
construct an orientifold using a $\mathbb{Z}_2$-symmetry of the bulk
action. On the fixed points we insert brane actions, which will turn out
to be invariant under the reduced ($N=1$) supersymmetry. For the moment
we will not restrict to domain walls (in this case eight-branes) since
our brane analysis is similar for orientifolds of lower dimension. In the
previous section we have seen that our bulk action possesses a number of
symmetries, among which a parity operation. To construct an orientifold,
the relevant $\mathbb{Z}_2$-symmetry must contain parity operations in
the transverse directions. Furthermore, in order to construct a charged
domain wall, we want for a $p$-brane the $(p+1)$-form R-R potential to be
even. For the $8$-brane the $I_9\Omega$ symmetry satisfies the desired
properties. For the other $p$-branes, it would seem natural to use the
$\mathbb{Z}_2$-symmetry
\begin{equation}
  I_{9,8,\ldots ,p+1} \Omega\equiv (I_9\Omega )(I_8\Omega )\cdots (I_{p+1}\Omega
  )\,,
 \label{defI98p}
\end{equation}
where $I_q\Omega $ is the transformation (\ref{I9Omega}) with $9$ replaced
by $q$, and $I_q$ and $\Omega $ commute. However, for some $p$-branes
($p=2,3,6,7$) the corresponding $C^{(p+1)}$ R-R-potential is odd under
this $\mathbb{Z}_2$-symmetry. To obtain the correct parity one must
include an extra $(-)^{F_L}$ transformation in these cases, which also
follows from T-duality~\cite{Dab98}, see appendix~\ref{app:Tduality}.
This leads for each $p$-brane to the $\mathbb{Z}_2$-symmetry indicated in
Table~\ref{tbl:Z2Op}.

\begin{table}[htb]
\begin{center}
\begin{tabular}{||c|c|c||}
\hline \rule[-1mm]{0mm}{6mm}
$p$       & IIB    & IIA  \\
\hline \rule[-1mm]{0mm}{6mm}
$9$      & $\Omega$   & - \\
\rule[-1mm]{0mm}{6mm}
$8$&-&$I_9\Omega$\\
\rule[-1mm]{0mm}{6mm}
$7$&$(-)^{F_L}I_{9,8}\Omega$&-\\
\rule[-1mm]{0mm}{6mm}
$6$&-&$(-)^{F_L}I_{9,8,7}\Omega $\\
\rule[-1mm]{0mm}{6mm}
$5$&$I_{9,8,\ldots ,6}\Omega$&-\\
\rule[-1mm]{0mm}{6mm}
$4$&-&$I_{9,8,\ldots ,5}\Omega$\\
\rule[-1mm]{0mm}{6mm}
$3$&$(-)^{F_L}I_{9,8,\ldots ,4}\Omega$&-\\
\rule[-1mm]{0mm}{6mm}
$2$&-&$(-)^{F_L}I_{9,8,\ldots ,3}\Omega$\\
\rule[-1mm]{0mm}{6mm}
$1$&$I_{9,8,\ldots ,2}\Omega$&-\\
\rule[-1mm]{0mm}{6mm}
$0$&-&$I_{9,8,\ldots ,1}\Omega$\\
\hline
\end{tabular}
\caption{The $\mathbb{Z}_2$-symmetries used in the orientifold
construction of an O$p$-plane. The T-duality transformation from IIA to
IIB in the lower dimension induces each time a $(-)^{F_L}$. }
\label{tbl:Z2Op}
\end{center}
\end{table}

Thus the correct $\mathbb{Z}_2$-symmetry for a general IIA O$p$-plane is
given by
\begin{equation}
((-)^{F_L})^{p/2} I_{9,8,\ldots ,p+1}\Omega\, . \label{z2}
\end{equation}
The effect of this $\mathbb{Z}_2$-symmetry on the bulk fields reads (the
underlined indices refer to the worldvolume directions, i.e.~$\mu =
(\underline{\mu}, p+1,\ldots ,9)$
\begin{align}
  \big\{ x^{p+1}, \ldots, x^9 \big\}
  & \rightarrow -
  \big\{ x^{p+1}, \ldots, x^9 \big\} \, , \notag \\
  \big\{ \phi, g_{\underline{\mu} \nuv}, B_{\underline{\mu} \nuv} \big\}
  & \rightarrow
  \big\{ \phi, g_{\underline{\mu} \nuv}, -B_{\underline{\mu} \nuv} \big\}\,, \notag \\
  \big\{ \A{5}_{\underline{\mu_1} \cdots \underline{\mu_5}},
    \A{9}_{\underline{\mu_1} \cdots \underline{\mu_9}},
    \G{2}_{\underline{\mu }\nuv}
    \big\}
  & \rightarrow (-)^{\tfrac{p}{2}}
    \big\{
    \A{5}_{\underline{\mu_1} \cdots \underline{\mu_5}},
    \A{9}_{\underline{\mu_1} \cdots \underline{\mu_9}},
    \G{2}_{\underline{\mu }\nuv}
    \big\}\, , \notag \\
  \big\{
    \A{7}_{\underline{\mu_1} \cdots \underline{\mu_7}},
    \G{0}, \G{4}_{\underline{\mu _1}\cdots \underline{\mu _4}}\big\}
  & \rightarrow (-)^{\tfrac{p}{2}+1}
    \big\{
    \A{7}_{\underline{\mu_1} \cdots \underline{\mu_7}},
    \G{0}, \G{4}_{\underline{\mu _1}\cdots \underline{\mu _4}}\big\}\, ,
    \displaybreak[2] \notag \\
  \big\{ \psi_{\underline{\mu}}, \epsilon \big\}
  & \rightarrow
    -\alpha \Gamma^{p+1 \cdots 9} (-\Gamma_{11})^{\tfrac{p}{2}}
    \big\{ \psi_{\underline{\mu}},
    \epsilon \big\}\, , \notag \\
  \big\{ \lambda \big\}
  & \rightarrow + \alpha \Gamma^{p+1 \cdots 9} (+\Gamma_{11})^{\tfrac{p}{2}}
    \big\{ \lambda \big\}\,,
\label{IIAsymmetry}
\end{align}
and for fields with other indices there is an extra minus sign for each
replacement of a worldvolume index $\underline{\mu }$ by an index in a
transverse direction. We have left open the possibility of combining the
symmetry with the sign change of all fermions. This possibility
introduces a number $\alpha =\pm 1$ in the above rules. This symmetry
will be used for the orientifold construction.
\par
For this purpose we choose spacetime to be $\mathcal{M}^{p+1} \times
{T}^{9-p}$ with radii $R^{\overline{\mu}}$ of the torus that may depend
on the world-volume coordinates. All fields satisfy
\begin{equation}
\Phi(x^{\overline{\mu}})=\Phi(x^{\overline{\mu}}+2\pi
R^{\overline{\mu}})\, ,\label{cyclicfields}
\end{equation}
with $\overline{\mu}=(p+1,\ldots,9)$. We only keep fields that are even
under the appropriate parity symmetry (\ref{z2}). In the bulk this relates
fields at $x^{\overline{\mu}}$ and $-x^{\overline{\mu}}$. At the fixed
point of the orientifolds, however, this relation is local and projects
out half the fields. This means that we are left with only $N=1$
supersymmetry on the fixed points, where the branes will be inserted.
Consider for example a nine-dimensional orientifold. The projection
truncates our bulk $N=2$ supersymmetry to $N=1$ on the brane; only half
of the 32 components of $\epsilon$ are even under (\ref{IIAsymmetry}).
The original field content, a $D=10$, $(128 + 128)$, $N=2$ supergravity
multiplet, gets truncated on the brane to a reducible $D=9$, $(64 + 64)$,
$N=1$ theory consisting of a supergravity plus a vector multiplet. One
may further restrict to a constant torus. This particular choice of
spacetime then projects out a $N=1$ $(8+8)$ vector multiplet (containing
$e_{\dot{9}}{}^9$), leaving us with the irreducible $D=9$, $(56 + 56)$,
$N=1$ supergravity multiplet. Similar truncations are possible in lower
dimensional orientifolds, on which the $(64 + 64)$ $N=1$ theory also
consists of a number of multiplets.
\par
We propose the $p$-brane action ($p=0,2,4,6,8$) to be proportional to
\begin{align}
  \mathcal{L}_p = -e^{-\phi} \sqrt{-g_{(p+1)}}
 - \alpha \tfrac{1}{(p+1)!} \varepsilon^{(p+1)} \C{p+1} \,,
\text{~~with~} \varepsilon^{(p+1)} \C{p+1} \equiv
  \varepsilon^{(p+1)}_{\underline{\mu_0} \cdots \underline{\mu_p}} \,
  \C{p+1}{}^{\underline{\mu_0} \cdots \underline{\mu_p}} \,,
\label{braneaction}
\end{align}
with $\varepsilon^{(p+1) \; \underline{\mu_0} \cdots \underline{\mu_p}} =
 \varepsilon^{(10) \; \underline{\mu_0} \cdots \underline{\mu_p}
 \dot{p+1} \cdots \dot{9}}$, which follows from
$e_{\underline{\mu}}{}^{\overline{a}}=0$ (being odd). Here the underlined
indices are $(p+1)$-dimensional and refer to the world-volume. The
parameter $\alpha$ is the same that appears in (\ref{IIAsymmetry}) and
takes the values $\alpha=+1$ for {\it branes}, which are defined to have
tension and charge with the same sign in our conventions, and $\alpha=-1$
for {\it anti-branes}, which are defined to have tension and charge of
opposite signs.  Note that due to the vanishing of $B$ on the brane the
potentials $\C{p+1}$ and $\A{p+1}$ are equal. The $p$-brane action can
easily be shown to be invariant under the appropriate $N=1$ supersymmetry:
\begin{align}
  \delta_\epsilon \mathcal{L}_p =
    - e^{-\phi} \sqrt{-g_{(p+1)}} \,
    \bar{\epsilon} \big( 1- \alpha \Gamma^{p+1 \cdots 9}
    (\Gamma_{11})^{\tfrac{p}{2}} \big) \, \Gamma^{\underline{\mu}} \,
    \big( \psi_{\underline{\mu}} - \tfrac{1}{18} \Gamma_{\underline{\mu}}
    \lambda \big) \, .
\label{delLp}
\end{align}
The above variation vanishes due to the projection under
(\ref{IIAsymmetry}) that selects branes or anti-branes depending on the
sign of $\alpha$ ($+1$ or $-1$ respectively).  In the following
discussions we will assume $\alpha=1$ but the other case just amounts to
replacing branes by anti-branes.
\par
By truncating our theory we are able to construct a brane action that only
consists of bosons and yet is separately supersymmetric. Having these at
our disposal, we can introduce source terms for the various potentials. In
general there are $2^{9-p}$ fixed points. The compactness of the
transverse space implies that the total charge must vanish. Thus the
total action will read
\begin{align}
  \mathcal{L} = & \mathcal{L}_{\text{bulk}} + k_p \mathcal{L}_p \Delta_p \,,
    \notag \\
  \text{with~~} \Delta_p \equiv &
  \big(\delta(x^{p+1}) - \delta(x^{p+1}- \pi R^{p+1})\big) \cdots
  \big(\delta(x^{9}) - \delta(x^{9}- \pi R^{9})\big)
\label{bulkplbraneaction}
\end{align}
where the branes at all fixed points have a tension and a charge
proportional to $\pm k_p$, a parameter of dimension
$1/\text{[length]}^{p+1}$. Since anti-branes do not satisfy the
supersymmetry condition \eqref{delLp}, we need both positive and negative
tension branes to accomplish vanishing total charge. As explained in the
introduction we are going to interpret the negative tension branes as
O-planes.

The equations of motion following from \eqref{bulkplbraneaction} induce a
$\delta$-function in the Bianchi identity of the $8-p$-form field
strength. In general, an elegant solution is difficult to find, but in one
special case the situation simplifies. This is the eight-brane case and
will be discussed in the next section.

But first let us notice another possibility. With the above choice of
spacetime, $\mathcal{M}^{p+1} \times (T^{9-p} / \mathbb{Z}_2)$, one can
place the branes only at the fixed points without breaking all
supersymmetry. However, making the identification
\begin{align}
  x^{\overline{\mu}} \sim - x^{\overline{\mu}} \,,
\end{align}
the projection under \eqref{IIAsymmetry} would be local everywhere. The
odd fields would not only vanish on the fixed points but in all spacetime.
Thus the action \eqref{braneaction} would be invariant everywhere and the
branes would be allowed in between the fixed points. Note that not only
on the fixed points but also in the bulk, only $N=1$ supersymmetry would
survive the modding out of the $\mathbb{Z}_2$-symmetry. We will not
consider this choice of spacetime.

\section{Supersymmetry of the D$8$--O$8$ System}
\label{8branesusy}  Mimicking the set-up of the five-dimensional case, we
have now set the stage to add eight-brane actions. Replacing the mass
parameter by a field $\G{0}$ at the cost of a nine-form $\A{9}$, our bulk
action has two $\mathbb{Z}_2$-symmetries. The one involving parity will be
used to truncate our $N=2$ theory to $N=1$ on the fixed points. The
addition of brane actions will modify the equation of motion of the
nine-form: $\G{0}$ only has to be constant between the branes.
\par
First we choose our spacetime to be $\mathcal{M}^9 \times \mathcal{S}^1$.
All fields satisfy $\Phi(x^9)=\Phi(x^9+2 \pi R)$ with
$R=R(x^{\underline{\mu}})$ the radius of $\mathcal{S}^1$. Furthermore,
the fields can be split up in even and odd under $I_9\Omega$. Modding out
this $\mathbb{Z}_2$-symmetry the odd fields vanish on the fixed points
$x^9=0$ and $x^9= \pi R \equiv - \pi R$ of the orientifold, where we will
put the branes. Using the parity symmetry, and taking a constant radius
for the circle, the $D=10$, $(128 + 128)$, $N=2$ supergravity multiplet
gets truncated on the brane to a reducible $D=9$, $(56+56)$, $N=1$
supergravity (see the previous section).
\par
We start with the nine-dimensional $8$-brane  action placed at $x^9=0$
\begin{align}
  (S_{D8})_{x^9=0} =\mu _8 \int d^{10} x {\cal L}_8 \, \delta(x^9)
  = - \tau_{8} g_s \int d^{10} x \{e^{-\phi} \sqrt{-g_{(9)}}
  +\alpha  \tfrac{1}{9!} \varepsilon^{(9)} \C{9} \}\delta(x^9)\, ,
\label{8braneaction}
\end{align}
with $\varepsilon^{(9) \; \underline{\mu_0} \cdots \underline{\mu_8}}
\equiv \varepsilon^{(10) \; \underline{\mu_0} \cdots \underline{\mu_8}
\dot{9}}$ and $\sqrt{-g_{(9)}}=\sqrt{-g_{(10)}}$ where we
 use that $e_{\underline{\mu}}^{\; \; 9}=0$ (being odd).
The overall constant $\mu_8$ is related to the `physical' tension
$\tau_8$  as follows:
\begin{equation}
\mu_8= \tau_8 g_s = \frac{2 \pi}{(2\pi \ell_s)^9}\,. \label{tension}
\end{equation}
The underlined indices are nine-dimensional in this section.
\par
In this paper we assume that there is no matter on the branes. Thus, we
are describing the vacuum solution of the D-brane system, switching off
the excitations on the branes. As discussed in the Introduction, for the
total charge to vanish while maintaining supersymmetric equilibrium, one
needs negative tension branes rather than anti-branes.  We associate the
negative-tension branes to Orientifold(O)-planes. We suggest the
following action of the supersymmetric O$8$-plane at $x^9=0$:
\begin{align}
  (S_{O8})_{x^9=0} = 16 \tau_{8} g_s \int d^{10} x \{e^{-\phi} \sqrt{-g_{(9)}}
  +\alpha  \tfrac{1}{9!} \epsilon^{(9)} \C{9} \}\delta(x^9)\,.
\label{8planeaction}
\end{align}
Each plane has a charge $-16$  and thus we may associate this object with
16 negative tension D$8$-branes without matter. Both brane \& plane action
satisfy the supersymmetry condition (\ref{delLp}). We would like to
stress that supersymmetry of the brane action tells us to use always the
same value of $\alpha$, the relative sign between the kinetic and the WZ
term independently of the sign in front of the total brane action.
 Thus we are left with only D-branes and O-planes.

There are two ways to interpret {\it our complete effective action}. In
the first picture it consists of the bulk \& branes with positive and
negative tensions. In the alternative picture we have two O-planes as
well as a number of D-branes at each plane.
\par
i) {\it First picture: positive and negative tension branes.} In analogy
with the supersymmetric RS construction in $D=5$~\cite{Bergshoeff:2000zn}
we may first consider the complete action for  the IIA theory in the bulk
with $2k$ coincident positive tension branes placed at $x^9=0$ and $2k$
coincident negative tension branes placed at $x^9=\pi R$:
\begin{align}
S_1 =
 S_{\text{bulk}}
  +  2k (S_{D8})_{x^9=0} -2k (S_{D8})_{x^9=\pi R}\,.
  \label{bulk+D-D}
\end{align}
Here $k$ is an arbitrary integer.  We take $2k$ branes with positive
tension and $2k$ with negative tension to have a simple relation of this
picture with the orientifold construction where one counts branes and
their images, so that the total number is even.
\par
ii) {\it Alternative interpretation: planes \& branes.}
Following~\cite{Polchinski:1996df}, we consider O$8$-planes at the fixed
points $x^9=0$ and $x^9=\pi R=-\pi R$ and they carry the $8$-brane charge
$-16$ each. The theory of Type~IIA supergravity under orientifold
truncation would be inconsistent unless an $SO(32)$ gauge multiplet
appears in the theory. This means that between these O-planes we have to
place 32 D$8$-branes. This is the effective action of type ${\rm
I^\prime}$ string theory. It is T-dual to Type~I string theory, which is
obtained by modding the IIB theory with the $\mathbb{Z}_2$-symmetry
$\Omega$ \eqref{Omega}. This also explains the origin of the 32
D$8$-branes: the Type~I gauge group $SO(32)$ can be seen to come from 32
unoriented (spacetime filling) D$9$-branes and performing T-duality
yields the 32 D$8$-branes~\cite{Horava:1996qa}.

In general these D$8$-branes can move between the O$8$-planes. However,
we will only place them at the fixed points. At the point $x^9=0$ we have
an O$8$-plane which contributes $-16$ to $2k$ in (\ref{bulk+D-D}) and we
have there also a stack of $2n$ D$8$-branes, thus $2k= 2(n-8)$. At the
second fixed point of the orientifold we have $-16$ from the O$8$-plane
and $2(16-n)$ from the stack of  D$8$-branes so that $-2k=- 2[8 - (16-n)]
= -2 (n-8)$. This means that at $x^9=0$ for $n>8$ there will be an
effective action of the positive tension  branes. At $\pi R$   for $n>8$
there will be an effective action of the negative tension branes. The
total action is
\begin{align}
S_2 =
 S_{\text{bulk}}
  + (  S_{O8}+ 2n S_{D8})_{x^9=0} +  (S_{O8} + 2(16-n) S_{D8})_{x^9=\pi R}
  \label{bulk+D+O} \, .
\end{align}
The two actions are equal for the special choice $k=n-8$:
\begin{equation}
 S_1= S_2 = S_{\text{bulk} +O8+D8}\ :\qquad   k=n-8\,.
 \label{k=n-8}
\end{equation}
It also follows that if at $x^9=0$ we want to have the total tension from
branes \& planes positive, i.e.  $k>0$, there is a restriction $n\geq 8$
so that
\begin{equation}
 8 \leq n\leq 16\,.
\label{n}
\end{equation}

\section{8-Brane Solution and Killing Spinors}\label{8brane}

The total effective action \eqref{bulk+D+O} is given by the bulk action
and an O$8$-plane and $2n$ D$8$-branes at $x^9=0$ and an O$8$-plane and
$32-2n$ D$8$-branes at $\pi R$.  To analyze its equations of motion, we
will only keep the fields participating in D$8$--O$8$ dynamics: the
metric, the dilaton, and the $0$-form field strength and the $9$-form
potential. Thus, our starting point will be the following bulk, brane \&
plane supersymmetric action:
\begin{eqnarray}
S_{\text{bulk}+O8+D8}&=&
 \frac{2\pi}{(2\pi \ell_s)^8} \left [
{\displaystyle\int} d^{10}{x}\, -\sqrt{|{g}|} \left\{  e^{-2{\phi}} \left[
{R} -4\left( \partial{\phi} \right)^{2}\right] +{\textstyle\frac{1}{2}}
\left({G}^{(0)}\right)^{2} - {G}^{(0)} \star (d \C{9})
\right\}\right. \nonumber\\
&& \hspace{1.5cm} \left.-  \frac{2 (n-8)}{(2\pi \ell_s)} \{  e^{-\phi}
\sqrt{|g_{(9)}|}
 +\alpha \tfrac{1}{9!} \varepsilon^{(9)} \C{9} \}
  \left(\delta(x^9)- \delta(x^9- \pi R)\right)\right]. \label{bulk+8O+8D}
\label{O+D}
\end{eqnarray}
The D$8$-brane solution is given by
\begin{align}
\rmd {s}^{2}_{s} & \, =  H^{-1/2}_{D8}[-\rmd t^{2} + (\rmd
x^{\underline{\mu}})^2]
+ H^{1/2}_{D8} (\rmd x^9)^{2}\, ,\nonumber\\
e^{{\phi}} & \, = e^{{\phi}_{0}} H^{-5/4}_{D8}\, ,\nonumber\\
\G{0} & = \, \alpha e^{-\phi_0} \partial_{\dot 9} H_{D8}
  = \alpha \, {n-8 \over 2\pi \ell_s} \varepsilon(x^9)
    \nonumber \\
  \C{9}_{\dot{0} \cdots \dot{8}}
  & = \, \alpha\, e^{-\phi_0} \left(H^{-1}_{D8}-1 \right)\,,
  \text{~~~~with~~}
  H_{D8} =1 -h_{D8}|x^9|\,, \qquad
  h_{D8}= {\displaystyle\frac{(n-8) g_s}{2\pi\ell_{s}}}\,,
\label{8solution}
\end{align}
where the first term of $H_{D8}$ is fixed by requiring $e^{\phi} =
e^{\phi_0}$ at $x^9=0$. This constant can be identified with the string
coupling constant $g_{s}$. This is a natural identification in the
non-asymptotically flat spacetimes associated to the higher branes and is
also consistent, via T-duality, with the standard identification of
$g_{s}=e^{\phi_{0}}$ where now this is the value of the dilaton at
infinity in the asymptotically flat spacetimes associated to
lower-dimensional branes.

For the non-vanishing fields of the D$8$-brane solution the Killing
spinor equations take the form
\begin{align}
\left( \partial_{{\mu}}+\tfrac{1}{4} \not\!{\omega}_{{\mu}} +\tfrac{1}{8}
e^\phi {G}^{(0)} {\Gamma}_{{\mu}} \right){\epsilon} = 0 \,, \qquad
\left(\not\!\partial\phi  +\ft{5}{4} e^\phi {G}^{(0)} \right) {\epsilon}
= 0 \,.
\end{align}
The ingredients of these equations are the non-vanishing zehnbeins and
spin connection components, the dilaton and ${G}^{(0)}$ . These can be
read off from \eqref{8solution} while the spin connection takes the form
\begin{align}
  \not\! \omega_{\underline{\mu}}
  = 2 {\Gamma}^9 {\Gamma}_{\underline{\mu}}
  \partial_{\dot9} H_{D8}^{-1/4} \,, \qquad
  \not\! \omega_{\dot 9} =0 \,.
\end{align}
For these fields the Killing spinor equations are solved by
\begin{align}
  {\epsilon} & \, = H^{-1/8}{\epsilon}_{0} \,,
  \text{~~with~}
  (1 + \alpha {\Gamma}^{9}) \, {\epsilon}_{0} =0 \,,
\end{align}
where ${\epsilon}_{0}$ is a constant spinor that satisfies the above
linear constraint. Thus it has 1/2 of unbroken supersymmetry of Type~IIA
theory, i.e. 16 unbroken supersymmetries.

\section{Critical Distances and Quantization of Mass } \label{ss:quantMcosmo}
{}From the $8$-brane solution (\ref{8solution}) one can read off that
$H_{D8}$ is zero for $|x^9|=1/h_{D8}$, implying singularities. Thus the
distance between the branes must be less than $1/h_{D8}$ so that the
harmonic function does not vanish. The radius of the circle and distance
between the O-planes is therefore restricted to
\begin{equation}
  R < {2\pi \ell_s\over (n-8) g_s}\,.
\end{equation}
$1/h_{D8}$ is called the critical distance. Thus it seems that type ${\rm
I^\prime}$ supergravity is consistent only on $\mathcal{M}^9 \times
(\mathcal{S}^1 / \mathbb{Z}_2)$ with a circle of restricted radius. Of
course we have only considered a special case of the type ${\rm I^\prime}$
theory with all D-branes on one of the fixed points. However, also with
D-branes in between the O-planes we expect the vacuum solution to imply a
critical distance. The same phenomenon of Type ${\rm I^\prime}$ was found
in \cite{Polchinski:1996df} in the context of the duality between the
Heterotic and Type I theories. Note that the maximal distance depends on
the distribution of the D-branes. In the most asymmetric case ($n=16$) it
is smallest while in the most symmetric case ($n=8$) there is no
restriction on $R$.

The $8$-brane solution \eqref{solution} has other consequences as well.
The equation of motion of the nine-form is modified by the brane \& plane
actions and leads to
\begin{equation} \G{0} = \alpha\, \frac{n-8}{2\pi \ell_s}
\varepsilon(x^9)\,. \label{G0}
\end{equation}
Thus we may identify the mass parameter of Type~IIA supergravity as
follows:
\begin{equation}
m= \left\{
  \begin{array}{cc}
\alpha {\displaystyle\frac{n-8}{2\pi\ell_{s}}}\, ,\,\,\,\, & x^9 > 0\, ,\\
& \\
-\alpha {\displaystyle\frac{n-8}{2\pi\ell_{s}}}\, ,\,\,\,\, & x^9 < 0\, .\\
  \end{array}
\right. \label{quant}
\end{equation}
The mass is quantized in string units and it is proportional to $n-8$
where there are $2n$ and $2(16-n)$ D$8$-branes at each O$8$-plane. The
mass vanishes only in the special case $n=8$ when the contribution from
the D$8$-branes cancels exactly the contribution from the O$8$-planes. In
general, due to restriction (\ref{n}) the mass      takes only the
restricted values
\begin{equation}
2\pi\ell_{s} |m| = 0,\, 1,\, 2,\, 3,\, 4,\, 5,\, 6,\, 7,\, 8 .
\label{discrete}
\end{equation}
This is a quantization of our mass parameter, and for the cosmological
constant it follows that
\begin{equation}
m^2= (\G{0})^2 =  \left (\frac{n-8} {2\pi \ell_s} \right )^2\,. \label{CC}
\end{equation}
Thus the mass parameter and the cosmological constant are quantized in
the units of the string length in terms of the integers $n-8$.
\par
The quantization of the mass and of the cosmological constant in $D=10$
was discussed before
in~\cite{Polchinski:1995mt,Polchinski:1996df,Polchinski:1996sm} as well as
in~\cite{Bergshoeff:1996ui,Green:1996bh}.  In the latter two references,
two independent derivations of the quantization condition were given.
In~\cite{Bergshoeff:1996ui}, the T-duality between a $7$-brane \&
$8$-brane solution was investigated. Here it was pointed out that, in the
presence of a cosmological constant, the relation between the $D=10$ IIB
R-R scalar $\C{0}$ and the one reduced to $D=9$, $c^{(0)}$, is given via
a generalized Scherk--Schwarz prescription:
\begin{equation}
\C{0} = c^{(0)} (x^9) + m x^8 \,. \label{axion}
\end{equation}
Here $(x^8, x^9)$ parametrize the 2-dimensional space transverse to the
7-brane. $x^9$ is a radial coordinate whereas $x^8$ is periodically
identified (it corresponds to a U(1) Killing vector field):
\begin{equation}
x^8 \sim x^8 +1 \, . \label{isometry}
\end{equation}
Furthermore, due to the $SL(2,\mathbb{Z})$ U-duality, the R-R scalar
$\C{0}$ is also periodically identified:
\begin{equation}
\C{0} \sim \C{0} +1 \,.
\end{equation}
Combining the two identifications with the reduction rule for $\C{0}$
leads to a quantization condition for $m$ of the form
\begin{equation}
m \sim \frac{n}{\ell_s}\, ,\hskip 2truecm n\,\,\,{\rm integer}\, .
\end{equation}
The same result was obtained by a different method in \cite{Green:1996bh}.

We are able to give a new, and independent, derivation of the
quantization condition for the mass and cosmological constant. The
conditions given in (\ref{quant}), (\ref{CC}) follow straightforwardly
from our construction of the bulk \& brane \& plane action
(\ref{bulk+8O+8D}).

Note that the Scherk--Schwarz reduction in (\ref{axion}) and the
quantization of $SL(2,\mathbb{R})$ were essential in deriving the quantization of
$m$. In the new dual formulation we can derive a similar T-duality
relation between the 7-brane and the 8-brane, including the source
terms.  However, in this case the T-duality relation does not imply a
quantization condition for $m$ since we do not know how to realize the
$SL(2,\mathbb{R})$ symmetry in the dual formulation. Another noteworthy feature is
that the derivation of the T-duality rules in the dual formulation does
not require a Scherk-Schwarz reduction. This is possible due to the fact
that the R-R scalar only appears after solving the equations of motion.

\section{BPS Action and Supersymmetric Flow Equations} \label{ss:BPSact}
Following our work in $D=5$, we consider the supersymmetric flow
equations corresponding to a domain wall solution. Originally,
supersymmetric flow equations were introduced in the context of black
holes in~\cite{Ferrara:1997tw}. They follow from the BPS-type energy
functional, which has a form of a sum of perfect squares, up to total
derivatives. Supersymmetric flow equations are simply the requirement
that each expression in the perfect square vanishes. Typically it is the
same condition that may be derived from the Killing spinor equations or
from the field equations.
\par
For the domain walls, the corresponding BPS-type energy functional was
derived in~\cite{Skenderis:1999mm,DeWolfe:1999cp,Kallosh:2000tj}. It
consists usually of two perfect squares, one with a positive and one with
a negative sign and some total derivative terms. The requirement that
each expression in the perfect square vanishes, leads to the
supersymmetric flow equations. However, for the domain walls the presence
of kinks requires a more careful treatment of the jump conditions at the
wall, as shown in~\cite{Bergshoeff:2000zn} for the 3-branes in $D=5$. With
supersymmetric bulk \& brane \& plane actions this will be taken care of
automatically as we will show below.
\par
We start with the  action (\ref{bulk+8O+8D}) and look for configuration
which depends only on $x^9$. We  choose the metric in the form suitable
for D$p$-branes in general.
\begin{equation}
  \rmd s^2 = f^2(x^9) \, (-\rmd t^2+(\rmd x^{\underline{\mu}})^2) + f^{-2}(x^9) \, (\rmd x^9)^2 \,.
\label{Dpbrane}
\end{equation}
The scalar curvature is (with prime indicating a derivative with respect
to $x^9$)
\begin{equation}
  R= 18\left [5 (f')^2 + f f^{''}\right ].
\label{curvDp}
\end{equation}
The expression for the energy functional consists of the contribution
from the bulk, and from the planes \& branes. The branes are  at the
positions of the O-planes.
\par
For our ansatz for the time-independent configuration, the energy is minus
the Lagrangian $E=-{\cal L}$. We find the following expression for the
energy functional of the bulk \& brane-plane actions with $k=n-8>0$
\begin{align}
\frac{(2\pi \ell_s)^8}{2\pi} E_{\rm total} = & \, \tfrac{1}{2} f^8
e^{-2\phi} (f \phi' - \tfrac{5}{4}  \alpha \G{0} e^{\phi} )^2-
  18 f^8 e^{-2\phi} (2 f'- \tfrac{1}{2}  f  \phi'
   + \tfrac{1}{8} \alpha \G{0} e^{\phi} )^2 + \nonumber\\
& + 18   (f^9 f' e^{-2\phi})'   +( \alpha\G{0} f^9 e^{-\phi})' \nonumber\\
& + \left({2k \over 2\pi \ell_s}(\delta(x^9) - \delta(x^9- \pi R)
 - \alpha
{\G{0}}'\right)( f^9 e^{-\phi}
+  \alpha \tfrac{1}{9!} \varepsilon^{(9)}
\C{9})\, . \label{totalE}
\end{align}
The third line in this expression can be removed by solving equations of
motion for the 9-form field which leads to
\begin{equation}
{2k \over 2\pi \ell_s} \left((\delta(x^9) - \delta(x^9-\pi R)\right) -
\alpha {\G{0}}'=0 \qquad
  \Longrightarrow \qquad
  \G{0}=\alpha {k \over 2\pi \ell_s}\varepsilon(x^9) \,,
\label{formequation}
\end{equation}
which is the same result as before. The second line of (\ref{totalE})
consists of total derivatives, which vanish in our space with fields
satisfying (\ref{cyclicfields}). Thus the final expression for the total
energy is given by
\begin{align}
{(2\pi \ell_s)^8 \over 2\pi} \oint E_{total} = \oint \tfrac{1}{2} f^8
e^{-2\phi} \big[ (f \phi' - \tfrac{5}{4}  \alpha \G{0} e^{\phi} )^2
  - 36 (2 f'- \tfrac{1}{2}  f  \phi' +  \tfrac{1}{8}
  \alpha \G{0} e^{\phi} )^2 \big] \,.
\label{final}
\end{align}
The difference with the structure of the BPS action in previous
cases~\cite{Skenderis:1999mm,DeWolfe:1999cp,Kallosh:2000tj}  is due to the
presence of the mixed $f'\phi'$ terms, which originate from the second
derivative of the metric converted into the first derivative of the
metric and the dilaton. In Einstein frame the procedure of getting rid of
the second derivative of the metric does not involve the dilaton
derivative.
\par
The supersymmetric flow equations are
\begin{equation}
  f{\phi}' = \tfrac{5}{4}  \alpha \G{0} e^{{\phi}}\, , \qquad 2 f'- \tfrac{1}{2}  f  \phi' =-
 \tfrac{1}{8} \alpha \G{0} e^{{\phi}}\, .
\label{susyflow}
\end{equation}
We may eliminate $\alpha \G{0}$ from these equations so that ${\phi}' = 5
\, f' / f$. If we choose $c f=  e^{{1\over 5}{\phi}}$ where $c$ is an
arbitrary positive\footnote{We focus here on the special case when the
energy of the brane at $x=0$ is positive. Note that the energy is given
by ${2k \over 2\pi \ell_s} f^9 e^{-{\phi}}$, thus $f>0$.} constant, we get
\begin{equation}
\left( e^{-{4\over 5}\phi}\right)'=  - c \alpha \G{0}\, , \qquad  \left(
e^{-{4\over 5}\phi}\right)^{''}=- c \alpha (\G{0})'= - c {2k \over 2\pi
\ell_s}\left(\delta(x^9)- \delta (x^9- \pi R)\right)\, . \label{susyflow2}
\end{equation}
We may choose two constants in our solution to be defined by the initial
values at $x^9=0$ of the metric and of the dilaton so that $f(0)=1$ and
$e^{\phi}(0) = g_s=c^5$. For such a choice, the solution is given in terms
of a harmonic function $H_{D8}$
\begin{equation}
H_{D8} \equiv   f^{-4} = c^4 e^{-{4\over 5}\phi}
  = 1- g_s {k \over 2\pi \ell_s}|x^9 |
  = 1- g_s |m||x^9|\, .
\label{solution}
\end{equation}
where $k=n-8>0$. Note that the first term in the harmonic function has to
be positive. At small $|x|$ the exponent has to be positive, and also we
should be able to take a square root of it to have $f^2 = H^{-1/2}$. The
second term comes out negative: this means that there is a singularity at
$H=0$. The string coupling $e^\phi$ blows up at $|x^9|=|x^9|_{\rm
critical}= 1 / (g_s \, |m|)$ and the metric is singular. Thus we have to
place the second wall at $\pi R< |x|_{\rm critical}$ .
\par
In the solutions of the second order equations we have also  found that
the two constants in the harmonic function are of opposite sign. This
supports the present picture that one needs two O-planes with branes on
them at the finite distance from each other, to describe a physically
meaningful configuration for supersymmetric domain walls (8-branes). It
is plausible that by adding some more non-vanishing fluxes in our
configurations that break more supersymmetries, one can find a solution
where the runaway behaviour of the dilaton is replaced by the critical
point with the fixed value  of the dilaton. In such a solution one may
expect that  an infinite distance between orientifold planes will be
possible. One natural candidate for such solution is the embedding of the
FGPW type solution~\cite{FGPW} into our  $D=10$ bulk \& brane
construction.

\section{Summary of Results and Discussion} \label{ss:summary}
The main new results of this paper of general nature are the new
formulations of Type~II $D=10$ supergravity (section~\ref{ss:susybulk}).
For both Type~IIA and IIB theories, we constructed democratic bulk
theories with a unified treatment of all R-R potentials.  Due to the
doubling of R-R degrees of freedom one had to impose extra duality
constraints and thus a proper action was not possible. A so-called
pseudo-action, containing kinetic terms for all R-R potentials but
without Chern-Simons terms, was discussed. Furthermore, we have broken
the self-duality explicitly in the IIA case, allowing for a proper
action.  Instead of all R-R potentials only half of the $\C{p}$'s occur
in these theories. Both the standard ($p=1,3$) as well as the dual
($p=5,7,9$) formulations were discussed. Using these actions all bulk \&
brane systems can be described.

In section~\ref{branesusy} we have studied brane actions at the fixed
points of orientifolds. It turned out that, on the appropriate
orientifolds, all brane actions preserve half of the $N=2$ supersymmetry.
Either branes or anti-branes fulfilled this condition but not both. This
can also be understood from the point of view of supersymmetric
equilibrium of forces. Thus, in a compact space, to have vanishing charge,
both positive and negative tension (anti-)branes must be used. The latter
can be interpreted as orientifold planes while the first correspond to
Dirichlet branes. One particular case of this was studied in detail: the
eight-brane. Our supersymmetric bulk \& brane action gave us a
description of the D$8$--O$8$ system. These are the fundamental objects
in Type ${\rm I^\prime}$ string theory. By studying this explicit
construction we have found several interesting features.

We have carefully studied the issue of the supersymmetric D$8$--O$8$
sources and have found that the  jump conditions on the walls are
satisfied for our BPS solutions in presence of sources. This is a highly
non-trivial issue in view of the recent studies of the difficulties  to
include the brane sources in the uplifted  RS scenario
in~\cite{Cvetic:2000id}.  Our BPS 8-brane solution implied a maximal
distance between the two walls in order to avoid singularities. At
\begin{align}
  |x^9|_{\text{critical}} = {2\pi \ell_s \over (n-8) g_s}\, ,
\hspace{.5cm} n \neq 8\, ,
\end{align}
the harmonic function vanishes and thus the second wall has to be placed
before this. Note that the maximal distance depends on the positions of
the D-branes.  The $n=8$ case is special since the D8-branes are
symmetrically distributed and hence are cancelled by the O-planes
contributions.  Also the mass parameter is found to be quantized in
string units:
\begin{align}
  |m|= {n-8 \over 2\pi \ell_s}\, ,
\end{align}
(again for $n\neq 8$).  Note that it is proportional to the D-brane
distribution factor $n-8$ while the maximal distance was inversely
proportional. In fact, we have the simple relation
$|x^9|_{\text{critical}} = 1/(g_s \, |m|)$.  This clearly shows the
twofold effect of introducing source terms: they induce a non-zero value
of the mass parameter but also imply a maximal distance between the
walls.  Of course this is all special to the exotic 8-brane since its
corresponding field strength $\G{10}$ does not fall off at infinity.

The D$8$-brane configuration presented in this paper requires two domain
walls at the finite distance from each other. This is necessary to cut
off the singularity of the metric and to keep the dilaton from  blowing
up. On the other hand this solution also has too many unbroken
supersymmetries, (1/2 of all supersymmetries of Type~II string theories).
However, solutions with more non-vanishing form fields might exist with a
dilaton that does not blow up but tends to a fixed value and with a
metric given by the product of the $AdS_5$ space and some Euclidean
5-dimensional manifold. Our D$8$-branes can be wrapped around this 5d
manifold and produce the 3-branes in 5d space with Minkowski signature.
These kind of 3-branes have been used in the bulk \& brane construction
of~\cite{Bergshoeff:2000zn}. The distance between orientifold planes would
not be limited in such a situation and in particular, the second wall at
$\pi  R$ may be pushed to infinity. The number of unbroken supersymmetries
in the bulk would be equal to 1/4 or 1/8 of original 32 and on the wall we
would have the desirable $D=4$, $N=1$ supersymmetry. We expect the
relevant bulk solution to be the uplifted  FGPW solution~\cite{FGPW}
which has one IR fixed point at $|\tilde x^9|\rightarrow \infty$. The  UV
fixed point would be cut off by our O$8$-D$8$ plane at $x^9=0$. Due to
the presence of the D$8$ branes and O$8$ planes such solution would
realize the 5d RSII scenario in the framework of fundamental objects of
string theory.

A notable difference of our scenario from the HW~\cite{HW,StelleandCo}
scenario is that the walls are the  O$8$ and D$8$ objects which exist in
string theory. They may be wrapped around some $D=5$ manifold. The main
goal of the HW theory was to present a scenario for appearance of chiral
fermions starting with $D=11$ supersymmetric theory with non-chiral
fermions. Our O$8$-D$8$ construction may reach this precise goal in an
interesting and controllable way due to stringy nature of this
construction and due to the complete control over supersymmetries in the
bulk \& on the walls. We remark that the strong coupling limit of
Type~I$^\prime$ string theory is equal to the HW theory. Using the
results of this paper, it would be interesting to investigate whether and
how in this limit the O$8$-D$8$ objects can be related to the HW branes.


It is clear that the D8-O8 system can be generalized much further. To
start with, placing D-branes in any compact transverse space requires the
presence of oppositely charged branes that will have to have opposite
tensions in order to be in supersymmetric equilibrium. If all the
negative-tension branes are identified with orientifold planes, as we are
suggesting here that should be done, then the compact transverse spaces
must be orbifolds with the orientifold planes placed in the orbifold
points. The $\mathbb{Z}_{2}$ reflection symmetries associated to the
orientifold planes can be part of more general orbifold groups
($\mathbb{Z}_{n}$ etc.).  It would be interesting to realize these bulk
\& brane configurations explicitly.



\section*{Acknowledgments}

The authors would like to thank for useful discussions G.W.~Gibbons,
T.~Mohaupt, S.~Kachru and E.~Silverstein.

This work was supported by the European Commission RTN program
HPRN-CT-2000-00131, in which E.B.  is associated with Utrecht University.
The work of R.K.  was supported by NSF grant PHY-9870115. The work of
T.O.~has been supported in part by the Spanish grant FPA2000-1584.
T.O.~wouldlike to thank the C.E.R.N.~TH Division and the I.T.P.~of the
University of Groningen for their financial support and warm hospitality.
E.B.~would like to thank the I.F.T.-U.A.M./C.S.I.C.~for its hospitality.

\appendix
\section{Conventions}
\label{app:Conventions}
\subsection{General}
We use mostly plus signature $(-+\cdots +)$. Greek indices
$\mu,\nu,\rho\ldots$ denote world coordinates and Latin indices
$a,b,c\ldots$ represent tangent spacetime. They are related by the
vielbeins $e_{a}{}^{\mu}$ and inverse vielbeins $e_{\mu}{}^{a}$. Explicit
indices $0, \ldots, 9$ are dotted for world coordinates and undotted in
the tangent spacetime case. The covariant derivative (with respect to
general coordinate and local Lorentz transformations) is denoted by
$\nabla_\mu$. Acting on tensors $\xi$ and spinors $\chi$ it reads
\begin{align}
  \nabla_\mu \xi = & \, \partial_\mu \xi\, , \notag \\
  \nabla_\mu \xi^\nu = & \, \partial_\mu \xi^\nu +
\Gamma_{\mu \rho}^{\; \; \; \; \nu} \xi^\rho\, , \notag \\
  \nabla_\mu \chi = & \, \partial_\mu \chi +
\tfrac{1}{4} \omega_{\mu}^{\; \; ab} \Gamma_{ab} \chi\, , \notag \\
  \nabla_\mu \chi^\nu = & \, \partial_\mu \chi^\nu +
\Gamma_{\mu \rho}^{\; \; \; \; \nu} \chi^\rho +\tfrac{1}{4}
\omega_{\mu}^{\; \; ab} \Gamma_{ab} \chi\, .
\end{align}
Here $\Gamma_{\mu \rho}^{\; \; \; \; \nu}$ and $\omega_{\mu}^{\; \; ab}$
are the affine and spin connection, respectively. Indices with an
additional underlining indicate lower-dimensional brane indices. We
symmetrize and antisymmetrize with weight one. Slashes are also used in
the following sense: $\not\!\!H=H^{\mu \nu \rho }\Gamma _{\mu \nu \rho
}$, and $\not\!\!H_\mu =H_{\mu \nu \rho }\Gamma ^{\nu \rho }$.

Our conventions in form notation are as follows:
\begin{align}
  P^{(p)} \equiv & \, \frac{1}{p!} P^{(p)}_{\mu_1 \cdots \mu_p} \rmd x^{\mu_1} \wdg \cdots \wdg
     \rmd x^{\mu_p} \,, \notag \\
  P^{(p)}  \cdot Q^{(p)} \equiv & \, \frac{1}{p!}
  P^{(p)}_{\mu_1 \cdots \mu_p} Q^{(p)\,\mu_1 \cdots \mu_p}\,,
  \notag \\
  P^{(p)} \wdg Q^{(q)} \equiv
  & \, \frac{1}{p!q!} P^{(p)}_{\mu_1 \cdots \mu_p}
    Q^{(q)}_{\mu_{p+1} \cdots \mu_{p+q}} \rmd x^{\mu_1} \wdg \cdots \wdg
    \rmd x^{\mu_{p+q}} \,, \notag \\
  \star \, P^{(p)} \equiv & \, \frac{1}{(10-p)!p!} \sqrt{-g}
    \varepsilon^{(10)}_{\mu_1 \cdots \mu_{10}} P^{(p)\,\mu_{11-p} \cdots \mu_{10}}
    \rmd x^{\mu_1} \wdg \cdots \wdg \rmd x^{\mu_{10-p}} \,, \notag \\
  & \text{with~~} \, \varepsilon^{(10)}_{0123\cdots 9}=-\varepsilon ^{0123\ldots 9}=1 \,, \notag \\
  \star \star \, P^{(p)} = & \, (-)^{p+1} P^{(p)} \,, \notag \\
  d \equiv
  & \, \partial_\mu \rmd x^\mu \,,
\label{formconv}
\end{align}
where the last line is the exterior derivative, acting from the left.
Also we will use the following abbreviation:
\begin{align}
  {\bf e}^{\pm B} \equiv  \pm B + \tfrac{1}{2} B \wdg B
    \pm \tfrac{1}{3!} B \wdg B \wdg B + \ldots
\end{align}
\subsection{Spinors in Ten Dimensions}
The ten-dimensional $\Gamma$-matrices are defined to satisfy the
anticommutation relations
\begin{equation}
  \big\{ {{\Gamma}}{}^a, {\Gamma}{}^{{b}} \big\} =
  +2 \eta^{ab} \, .
\end{equation}
We can choose a Majorana representation where they are purely real, with
the choice ${{\cal C}} ={\Gamma}_0$ for the charge conjugation matrix.
Their Hermiticity properties are
\begin{equation}
  {{\Gamma}}^{0 \; \dagger} = \Gamma ^{0\; T}=-{{\Gamma}}^0\, , \qquad
  {{\Gamma}}^{i \; \dagger} =\Gamma ^{i\; T}={{\Gamma}}^i\, , \hspace{.5cm}
    i = 1,\ldots,9\, .
\end{equation}
Furthermore we have the useful $\Gamma$-matrices identity
\begin{align}
  & \Gamma_{11} \GAM{n}_{a_1 \cdots a_n} =
    \frac{(-1)^{\Int [(10-n)/2]+1}}{(10-n)!}
    \varepsilon^{(10)}_{a_1 \cdots a_n b_1 \cdots b_{10-n}}
    \Gamma^{(10-n) \; b_1 \cdots b_{10-n}} \, , \notag \\
  & \text{~~~ with ~~~}
    {\Gamma}^{(n)}_{{a}_1 \cdots {a}_n} \equiv
    {\Gamma}_{[ {a}_1} \cdots {\Gamma}_{{a}_n ]}
    \text{~~ and ~~} \Gamma_{11} \equiv  \Gamma^0 \cdots \Gamma^9 \, ,
\label{GammaDual}
\end{align}
which combines with the star operation to
\begin{equation}
  \Gamma _{11}{\Gamma}^{(n)}=(-)^{\Int\frac{n+1}{2}}\,\star \Gamma
  ^{(10-n)}\,.
 \label{starGamma}
\end{equation}

 The Majorana condition is equivalent to requiring all components of
a Majorana spinor to be real. We do not change order of the fermions in
performing complex conjugation. Using the above properties and the
definition of Majorana spinors one finds
\begin{equation}
  \overline{\chi}   \Gamma^{a_1\cdots a_n}   \psi =
  (-1)^{n+\Int\left[n/2\right]} \,   \bar\psi  \Gamma^{a_1\cdots a_n}
  \chi\, .
\end{equation}
\section{World-sheet T-duality} \label{app:Tduality}
Although the discussion on bulk \& brane supersymmetry has been in the
context of supergravity, it is also useful to consider the string origin.
Here we will derive a number of $\mathbb{Z}_2$-symmetries of Type~IIA and
IIB string theories, that are relevant for the orbifold\footnote{ An {\it
orbifold} corresponds to modding out a {\it spacetime}
$\mathbb{Z}_2$-symmetry. From the supergravity point of view this is the
correct terminology for section~\ref{ss:susybulk}. Here we will discuss
the string theory origin; a {\it world-sheet} operation is involved and
the corresponding construction is called an {\it orientifold}.
See~\cite{Dab98} for a general introduction. }
 procedure.
Since these are symmetries of the world-sheet, not only effective
supergravity actions but also the full string theories must be invariant
under the corresponding spacetime operations. It will turn out that they
are related via simple world-sheet T-duality relations. We will discuss
this in the Neveu--Schwarz--Ramond formalism. On the world-sheet one has
the following field content:
\begin{align}
  \big\{ X_L{}^\mu, X_R{}^\mu, \psi_L{}^\mu, \psi_R{}^\mu \big\}
\end{align}
with left- and right-moving bosons and fermions. To construct symmetries
of the two theories the following operations on these fields will be used:
\begin{center}$
\begin{array}{cccc}
  T_q:
  & \hspace{1cm} X_L{}^q \rightarrow -X_L{}^q \hspace{1cm}
  & \psi_L{}^q \rightarrow - \psi_L{}^q \\
  \\
  I_q:
  & X_L{}^q \rightarrow -X_L{}^q
  & \psi_L{}^q \rightarrow - \psi_L{}^q \\
  & X_R{}^q \rightarrow -X_R{}^q
  & \psi_R{}^q \rightarrow - \psi_R{}^q \\
  \\
  \Omega:
  & X_L{}^\mu  \leftrightarrow X_R{}^\mu
  & \psi_L{}^\mu \leftrightarrow \psi_R{}^\mu
  & \hspace{1cm} (\sigma \rightarrow -\sigma) \\
  \\
  (-)^{F_L^c}:
  & & \psi_L{}^\mu \rightarrow - \psi_L{}^\mu
\end{array}$
\end{center}
with the other fields invariant. Note that all these operations square to
one. Although the world-sheet action may be invariant under the above
operations, in string theory boundary conditions are involved as well.
Different choices of boundary conditions lead to different types of
string theories. Therefore, for an operation to be a symmetry of a
certain theory, it should also leave the boundary conditions invariant.
Bearing this in mind, it is easy to see that $\Omega$ is a symmetry of
Type~IIB theory only. Using T-duality between Type~IIA and IIB theory we
can derive other symmetries. This chain of symmetries, connected via
T-duality, reads
\begin{align}
  S_p \equiv ((-)^{F_L^c})^{\Int\big[ \tfrac{p}{2} \big]} \cdot I_9 \cdots I_{p+1} \Omega
    = T_{p} S_{p+1} T_{p} \,,
\end{align}
with $S_{10} \equiv \Omega$. It is explicitly given in
table~\ref{tbl:Z2Op}. Thus, by applying T-duality to a
$(p+1)$-dimensionally extended parity operation, we get a parity
operation with one more direction involved. The $S_p$ with $p$ odd are
symmetries of Type~IIA theory; the $S_p$ with $p$ even of Type~IIB theory.
Using these $Z_2$-symmetries it is possible to construct orientifolds of
any dimension that are charged under the corresponding R-R potential.
Furthermore, under T-duality the actions of orientifold planes of all
dimensions are related. This should not come as a surprise; an
orientifold plane can be seen as a truncation of a D-brane and the latter
are also related under T-duality.


\providecommand{\href}[2]{#2}\begingroup\raggedright\endgroup
\end{document}